\documentclass[a4paper,structabstract]{aa}  
\usepackage{graphicx}
\usepackage{float}
\usepackage{txfonts}
\usepackage{color}
\usepackage{natbib}
\bibpunct{(}{)}{;}{a}{}{,} % to follow the A&A style

\begin{document}
\newcommand{\pd}[2]{\frac{\partial #1}{\partial #2} }
\newcommand{\HALF}{\frac{1}{2}}
\newcommand{\DS}{\displaystyle}
\newcommand{\red}[1]{\color{red} #1 \color{black}}
\newcommand{\blue}[1]{\color{blue} #1 \color{black}}
\newcommand{\hvec}[1]{\hat{\vec{#1}}}
\newcommand{\E}{{\cal E}}
\newcommand{\av}[1]{\left<#1\right>}
%
%   \title{An Improved Orbital Advection Scheme for MHD in the PLUTO Code}
%   \title{Implementation of an Improved Orbital Advection Scheme for MHD
%          in the PLUTO Code}
%   \title{A Conservative Orbital Advection Scheme for Magnetized Sheared   Flows}
   \title{A conservative orbital advection scheme for simulations of magnetized shear flows with the PLUTO\thanks{Freely distributed at \texttt{http://plutocode.ph.unito.it}} code}

   \subtitle{}

   \author{A. Mignone\inst{1}, M. Flock\inst{2,3}, M. Stute\inst{4}, S. M. Kolb\inst{4} \and G. Muscianisi\inst{5}
          }

   \institute{Dipartimento di Fisica Generale, Universit\'a di Torino,
              via Pietro Giuria 1, 10125 Torino, Italy\\
              \email{mignone@ph.unito.it}
         \and
             CEA Irfu, SAP, Centre de Saclay, 91191 Gif-sur-Yvette, France
         \and 
              Max Planck Institute for Astronomy, K\"onigstuhl 17, 
             69117 Heidelberg, Germany\\
             \email{flock@mpia.de}
         \and
             Institute for Astronomy and Astrophysics,
             Section Computational Physics,
             Eberhard Karls Universit\"at T\"ubingen,
             Auf der Morgenstelle 10,
             D-72076 T\"ubingen,
             Germany\\
             \email{[matthias.stute,stefan.kolb]@tat.physik.uni-tuebingen.de}
         \and
              Consorzio Interuniversitario CINECA, via Magnanelli, 6/3, 40033 Casalecchio di Reno (Bologna), Italy}

%   \date{...}

% \abstract{}{}{}{}{} 
% 5 {} token are mandatory
 
  \abstract
{Explicit numerical computations of hypersonic or super-fast differentially rotating disks are subject to the time-step constraint imposed by the Courant condition, according to which waves cannot travel more than a fraction of a cell during a single time-step update.
When the bulk orbital velocity largely exceeds any other wave speed (e.g., sound or Alfv\'en), as computed in the rest frame, the time step is considerably reduced and an unusually large number of steps may be necessary to complete the computation.}
{We present a robust numerical scheme to overcome the Courant limitation by improving and extending the algorithm previously known as FARGO (Fast Advection in Rotating Gaseous Objects) to the equations of magnetohydrodynamics (MHD) using a more general formalism.
The proposed scheme conserves total angular momentum and energy to machine precision and works in Cartesian, cylindrical, or spherical coordinates.
The algorithm has been implemented in the next release of the PLUTO code for astrophysical gasdynamics and is suitable for local or global simulations of accretion or proto-planetary disk models.}
{By decomposing the total velocity into an average azimuthal contribution and a residual term, the algorithm approaches the solution of the MHD equations through two separate steps corresponding to a linear transport operator in the direction of orbital motion and a standard nonlinear solver applied to the MHD equations written in terms of the residual velocity.
Since the former step is not subject to any stability restriction, the Courant condition is computed only in terms of the residual velocity, leading to substantially larger time steps.
The magnetic field is advanced in time using the constrained transport method in order to fulfill the divergence-free condition.
Furthermore, conservation of total energy and angular momentum is enforced at the %discrete level by properly computing the source terms as combinations of flux-differences of upwind Godunov fluxes.
discrete level by properly expressing the source terms in terms of upwind Godunov fluxes available during the standard solver.
}
{Our results show that applications of the proposed orbital-advection scheme to problems of astrophysical relevance provides, at reduced numerical cost, equally accurate and less dissipative results than standard time-marching schemes.}
   {}


   \authorrunning{Mignone et al}
   \titlerunning{A FARGO Scheme for MHD}
   \maketitle
%
%________________________________________________________________

%%%%%%%%%%%%%%%%%%%%%%%%%%%%%%%%%%%%%%%%%%%%%%%%%%%%%%%%%%%%%%%%%%%%%%%%%%%%
\section{Introduction}
%
%
%
%
%%%%%%%%%%%%%%%%%%%%%%%%%%%%%%%%%%%%%%%%%%%%%%%%%%%%%%%%%%%%%%%%%%%%%%%%%%%%

The physics of accretion flows has received a great deal of attention over the past few decades, and several numerical investigations have contributed to broaden our current understanding of the problem in its diverse aspects.
Representative astrophysical scenarios involve magnetically driven turbulence in accretion disks around compact objects, which are typically modeled via a local approach \citep[the so-called shearing-box model of][]{HawGamBal.1995} or global disk simulations, that have recently become more appealing owing to increases  in the available computational power see, for instance, \citep[see, for instance][]{BecArmSim.2011, Flock_etal.2011, Sorathia_etal.2012}.
Likewise, in the context of planet formation, simulations of proto-planetary disks have become an important tool for the study of the dynamics of the gas and its impact on either particles or planets \citep[see][and references therein]{KBK09, UriKlaFloHen.2011}. %Flock 2011)

A common ingredient of all models is a highly supersonic or superfast, rotating, sheared flow.
In standard explicit numerical calculations, the time step is restricted, for stability reasons, by the Courant-Friedrichs-Lewy (CFL) condition \citep{CFL.1928}, which is formally stated as 
\begin{equation}\label{eq:CFL}
  \Delta t \sim C_a\frac{\Delta l}{\lambda_{\max}}\,,
\end{equation}
where $\Delta l$ is the cell length, $\lambda_{\max}$ is the fastest signal speed, and $C_a$ - the Courant number - is a limiting factor typically of $< 1$.
In essence, the CFL condition given in Eq. (\ref{eq:CFL}) prevents any wave from 
traveling more than a fraction of a grid cell.
This leads to a drastic reduction in the time step, whenever the orbital speed becomes considerably greater than either the sound or fast magnetosonic velocities, and results in excessively long computations. 
A typical occurrence is encountered in accretion or proto-planetary disks where the Keplerian velocity is the dominating flow speed. 

In the context of hydrodynamics, \citet{Masset.2000} presented the first fast Eulerian-transport algorithm for differentially rotating disks (FARGO) in which the timestep is limited by the velocity fluctuations around the mean orbital motion rather than by the total azimuthal velocity itself.
The algorithm is based on the simple recognition that if the orbital velocity is written as the sum of a constant average term plus a residual, the evolution operator can be implemented as the composition of a linear transport operator and a nonlinear step involving the residual velocity.
Since the former step yields a uniform shift along the orbital direction, the CFL condition is determined during the nonlinear step by the magnitude of the residual velocity.
The algorithm may be viewed as a temporary change to the local co-rotating frames placed at different radial positions of the differentially rotating disk.
The advantage of using such a strategy is twofold: on the one hand, it reduces numerical dissipation as the Keplerian flow is now treated as a mean flow in the equations and, on the other hand, it allows larger time steps to be taken thus reducing the computational costs (as long as the residual velocity remains subsonic).

The ideas and concepts behind FARGO were extended to magnetohydrodynamics (MHD) by \cite{JohGuaGam.2008} and \cite{StoGar.2010} in the context of shearing-box model which is a local approximation describing a small disk patch embodied by a Cartesian box co-rotating with the disk at some fiducial radius.
Because of the difficulties inherent in simulating an entire disk, local shearing-box models have largely contributed to much of what is presently known about the magneto-rotational instability and its implication for the transport of angular momentum in disks.
Nevertheless, several authors have stressed, over the past decade, the importance of global disk models, specially in the context of magneto-rotational instability, e.g., \citet{Armitage.1998}, \citet{Hawley.2000}, \citet{FroNel.2006} and \citet{RegUmu.2008}. 
Owing to the increased numerical power, high-resolution simulations of global disk models are now becoming amenable to investigation.
In this context, a first implementation of the FARGO scheme for MHD was presented by \cite{Sorathia_etal.2012} as part of the Athena code \citep{Stone_etal.2008} in cylindrical geometry.

Here we propose, using a somewhat different and more general formalism, an extension of the original FARGO scheme to the equations of MHD for different systems of coordinates in the context of Godunov-type schemes.
The algorithm differs from the original one in that it uses a fully unsplit integration scheme when solving the MHD equations in the residual velocity. 
Source terms arising from the velocity decomposition are treated carefully in order to restore the conservation of total energy and total angular momentum.
The magnetic field is evolved via the constrained-transport (CT) scheme, which preserves the divergence-free condition to machine accuracy at all times.
The proposed FARGO-MHD scheme is implemented in the PLUTO code for astrophysical gasdynamics \citep{PLUTO.2007, PLUTO.2011} and will be available with the next code release (4.0).
The algorithm has been fully parallelized so that domain decomposition can be performed in all three coordinate directions, thus allowing an efficient use of a large number of processors.

The paper is organized as follows. 
In section \ref{sec:method}, we briefly review the basic MHD equations and the time-step limitations imposed by standard explicit time-stepping methods (\S\ref{sec:std_approach}). 
The new FARGO-MHD scheme is presented in \S\ref{sec:FARGO} and consists of i) a standard nonlinear step (\S\ref{sec:sns}) formulated so as to preserves total energy and total angular-momentum conservation (\S\ref{sec:sns})
and ii) a linear transport step (\S\ref{sec:lts}) where fluid quantities are simply advected in the direction of orbital motion.
Numerical benchmarks and astrophysical applications are presented in Section \ref{sec:tests}, while conclusions are drawn in Section \ref{sec:summary}.

%%%%%%%%%%%%%%%%%%%%%%%%%%%%%%%%%%%%%%%%%%%%%%%%%%%%%%%%%%%%%%%%%%%%%%%%%
\section{Description of the FARGO-MHD scheme}
\label{sec:method}
%
%
%
%
%
%%%%%%%%%%%%%%%%%%%%%%%%%%%%%%%%%%%%%%%%%%%%%%%%%%%%%%%%%%%%%%%%%%%%%%%%%

%%%%%%%%%%%%%%%%%%%%%%%%%%%%%%%%%%%%%%%%%%%%%%%%%%%%%%%%%%%
\subsection{Standard approach and limitations}
\label{sec:std_approach}
%
%
%%%%%%%%%%%%%%%%%%%%%%%%%%%%%%%%%%%%%%%%%%%%%%%%%%%%%%%%%%%

We begin by considering the ideal MHD equations written as a nonlinear system of conservation laws
\begin{eqnarray}\label{eq:mhd_rho}
\DS \pd{\rho}{t} + \nabla\cdot\left(\rho\vec{v}\right) &=& 0 
 \\ \noalign{\medskip} \label{eq:mhd_mom}
\DS \pd{\vec{(\rho\vec{v})}}{t} + \nabla\cdot\left[
   \rho\vec{v}\vec{v}^T 
 -     \vec{B}\vec{B}^T\right] 
 + \nabla p_t&=& -\rho\nabla\Phi
\\ \noalign{\medskip}\label{eq:mhd_E}
\DS \pd{E}{t} + \nabla\cdot\left[
  \left(E + p_t\right)\vec{v} - 
  \left(\vec{v}\cdot\vec{B}\right)\vec{B}\right]& =& 
    -\rho\vec{v}\cdot\nabla\Phi
\\ \noalign{\medskip}\label{eq:mhd_B}
\DS \pd{\vec{B}}{t} - \nabla\times\left(\vec{v}\times\vec{B}\right) 
   &=& \vec{0}\,,
\end{eqnarray}
where $\rho$ is the mass density, $\vec{v}$ is the fluid velocity, $\Phi$ is the gravitational potential, $\vec{B}$ is the magnetic field, $p_t=p+\vec{B}^2/2$ is the total pressure accounting for thermal ($p$), and magnetic ($\vec{B}^2/2$) contributions. 
The total energy density $E$ is given by
\begin{equation}
 E = \frac{p}{\Gamma - 1} + \frac{1}{2}\rho\vec{v}^2
                                 + \frac{1}{2}\vec{B}^2\,,
\end{equation}
where an ideal equation of state (EoS) with specific heat ratio $\Gamma$ has been used.
Dissipative effects are neglected for the sake of simplicity, although they can be easily incorporated in this framework.

In the usual finite-volume approach, Eq. (\ref{eq:mhd_rho}) to Eq. (\ref{eq:mhd_B}) are discretized on a computational mesh spanned by the grid indices $i$, $j$, and $k$ corresponding to the location of a given cell in a three-dimensional (3D)  coordinate system.
We label the generic unit vector along a given axis with $\hvec{e}_d$ and consider, in what follows, Cartesian ($\{\hvec{e}_d\}=\hvec{x},\hvec{y},\hvec{z}$), cylindrical polar ($\{\hvec{e}_d\}=\hvec{R},\hvec{\phi},\hvec{z}$), and spherical ($\{\hvec{e}_d\}=\hvec{r},\hvec{\theta},\hvec{\phi}$) coordinates.
Individual cells have spatial extents given by $\Delta l_{d,ijk}$, where $d$ labels the direction.

The time step $\Delta t$ is determined by the CFL condition in Eq. (\ref{eq:CFL}), the precise form of which can depend on the employed time-stepping scheme.
In PLUTO, one can take advantage of either the corner-transport upwind (CTU) method of \citet{Colella.1990} and \citet{GarSto.2008} or resort to a Runge-Kutta (RK) discretization, where the spatial gradients are treated as the right-hand side of an ordinary differential equation (method of lines).
Both algorithms are dimensionally unsplit but stable under somewhat different Courant conditions.
Here we consider the 6-solve CTU and the second-order Runge-Kutta (RK2) scheme, which have the same number of Riemann problems per cell per step. 
If $N_{\rm dim}=2,3$ is the number of spatial dimensions, one has \citep{Beckers.1992, Toro.1999}
\begin{equation}\label{eq:Ca}
\left\{\begin{array}{lcll}
\DS  C_a = \Delta t \max_{ijk}\left[
           \max_d\left(\frac{|v_d| + c_{f,d}}{\Delta l_d}\right)\right] 
  & < & \DS \frac{1}{N_{\rm dim}-1}  &\quad\textrm{(CTU)}
\\ \noalign{\medskip}
\DS  C_a = \frac{\Delta t}{N_{\rm dim}}\max_{ijk}\left(
     \sum_d \frac{|v_d| + c_{f,d}}{\Delta l_d}\right)
    & < & \DS\frac{1}{N_{\rm dim}} &\quad\textrm{(RK2)} \quad,
\end{array}\right.
\end{equation}
where $v_d$ and $c_{f,d}$ are the fluid velocity and fast magnetosonic speed in the direction $d$.
For highly super-fast, grid-aligned flows, Eq. (\ref{eq:Ca}) shows that both CTU and RK2 yield comparable time steps in two-dimensions, while, in three-dimensions, RK2 is expected to give time steps that are approximately twice as large.

Nevertheless, numerical computations of advection-dominated flows for which $v_d \gg c_f$ may result in small time steps yielding unusually long computations and eventually suffering from an excess of numerical dissipation.
This kind of scenario worsens in the case of a Keplerian flow in polar or spherical coordinates, since the cell length becomes increasingly smaller towards the inner radius where the orbital flow velocity is faster.

%%%%%%%%%%%%%%%%%%%%%%%%%%%%%%%%%%%%%%%%%%%%%%%%%%%%%%%%%%%%%%%
\subsection{An orbital advection scheme for MHD}
\label{sec:FARGO}
%
%
%%%%%%%%%%%%%%%%%%%%%%%%%%%%%%%%%%%%%%%%%%%%%%%%%%%%%%%%%%%%%%%

To overcome the limitations outlined in the previous section, we now assume that the fluid velocity may be decomposed as $\vec{v}=\vec{v}'+\vec{w}$, where $\vec{w}$ is a solenoidal velocity field and $\vec{v}'$ is the fluctuation or residual.
Eqns. (\ref{eq:mhd_rho}) through (\ref{eq:mhd_B}) may then be written as
\begin{eqnarray}\label{eq:mhd1_rho}
\DS \pd{\rho}{t} + \nabla\cdot(\rho\vec{v}') + \vec{w}\cdot\nabla\rho
  &=&   0  
\\ \noalign{\medskip} \label{eq:mhd1_mom}
\DS  \pd{(\rho\vec{v}')}{t} 
   + \nabla\cdot\left(\rho\vec{v}'\vec{v}' - \vec{B}\vec{B} + \tens{I}p_t\right)
 + \vec{w}\cdot\nabla\left(\rho\vec{v}'\right) &=& 
\DS \vec{S}'_m - \rho\nabla\Phi
\\ \noalign{\medskip}\label{eq:mhd1_E}
\DS  \pd{E'}{t} 
  + \nabla\cdot\Big[(E' + p_t)\vec{v}' 
               - (\vec{v}'\cdot\vec{B})\vec{B}\Big]
   + \vec{w}\cdot\nabla E' &=& S'_E 
\\ \noalign{\medskip} \nonumber
                            & & \DS - \rho\vec{v}'\cdot\nabla\Phi
\\ \noalign{\medskip}\label{eq:mhd1_B}
\DS \pd{\vec{B}}{t} - \nabla\times\left(\vec{v}'\times\vec{B}\right)
                    - \nabla\times\left(\vec{w} \times\vec{B}\right)
  &=& 0     \,,
\end{eqnarray}
where $\tens{I}$ is the identity matrix and $E'$ is the residual energy density, defined as 
\begin{equation}
 E' = \frac{p}{\Gamma-1} + \frac{1}{2}\rho(\vec{v}')^2 
                         + \frac{1}{2}\vec{B}^2 \,,
\end{equation}
and the two additional source terms in the momentum and energy equations may be written, after some algebra, as
\begin{equation}\label{eq:Sm0}
 \vec{S}'_m = -\rho\vec{v}\cdot\nabla\vec{w} \,,
\end{equation}
and
\begin{equation}\label{eq:SE0}
 S'_E = - \rho\vec{v}'\cdot\left(\vec{v}\cdot\nabla\vec{w}\right)
      + \vec{B}\cdot(\vec{B}\cdot\nabla\vec{w})\,.
\end{equation}
Explicit expressions for these terms in different systems of coordinates can be found in Appendix \S\ref{sec:coord_eqns}.

We note that, Eqns (\ref{eq:mhd1_rho}) through (\ref{eq:mhd1_B}) have a general validity since no assumption has been made about the magnitude of the residual velocity $\vec{v}'$.
However, neither the residual momentum nor the residual energy density are conserved quantities because of the appearance of the additional source terms $\vec{S}'_m$ and $S'_E$ given by Eq. (\ref{eq:Sm0}) and (\ref{eq:SE0}).
Since nothing has of course changed at the continuous level, both the total energy and momentum must still be conserved.
The situation is different, however, at the discrete level where differential vector identities are satisfied only within the truncation error of the scheme. This is discussed in more detail in \S\ref{sec:source_terms}

If we look at the individual components of the system of Equations (\ref{eq:mhd1_rho}) through (\ref{eq:mhd1_B}), each evolution equation has the form
\begin{equation}\label{eq:model_eq}
 \pd{q}{t} + \nabla\cdot\vec{F}_q + \vec{w}\cdot\nabla q = S_q\,,
\end{equation}
where $q\in\left(\rho, \rho\vec{v}', E', \vec{B}\right)$, $\vec{F}_q$ is the usual MHD flux written in terms of the residual velocity $\vec{v}'$.
The source term $S_q$ appearing on the right-hand side of Eq. (\ref{eq:model_eq}) accounts for several contributions that include gravity, the additional term $\vec{S}'_m$ (or $S'_E$), and, in the case of curvilinear coordinate systems, geometrical factors arising from differential operators.
The last term on the left-hand side of Eq. (\ref{eq:model_eq}) describes the linear transport of $q$ with advection speed $\vec{w}$.
If $\vec{w}$ coincides with the average azimuthal velocity, this term has the simple effect of pushing fluid elements along their orbit.

An effective approach to solve Eq. (\ref{eq:model_eq}) is to use operator splitting and separate the solution into a standard nonlinear step that does not include the linear advection term $\vec{w}\cdot\nabla q$
%
%\begin{equation}\label{eq:LTS}
%  \pd{q}{t} + \vec{w}\cdot\nabla q = 0\,.
%\end{equation}
\begin{equation}\label{eq:SNS}
\DS {\cal L}_n(q):\quad \pd{q}{t} + \nabla\cdot\vec{F}_q = S_q \,,
\end{equation}
and a linear transport step corresponding to the solution of
\begin{equation}\label{eq:LTS}
\DS {\cal L}_t(q):\quad \pd{q}{t} + \vec{w}\cdot\nabla q = 0\,.
\end{equation}
The two operators are separately described in \S\ref{sec:sns} and \S\ref{sec:lts}, respectively.

When the orbital motion coincides with one of the coordinate axes, e.g. $w=\vec{w}\cdot\hvec{y}$, the exact solution of (\ref{eq:LTS}) can be formally expressed as $q(y,t^{n+1}) = q(y-w\Delta t^n, t^n)$, which corresponds to a non-integer translation of a row of computational zones. 
Since this operation is decomposed into an integer shift plus a remainder, the linear transport step is unconstrained and can be carried out for arbitrarily large time steps.

The time step is thus ultimately determined by the Courant condition during the nonlinear step, that is, by solving the regular MHD equations (plus additional source terms) written in terms of the residual velocity $\vec{v}'$.
Provided that $|v'_d| + c_{f,d} \ll |\vec{w}|$, the CFL condition in Eq. (\ref{eq:Ca}) now results in appreciably larger time steps, since the dominant background orbital contribution has now been removed from the computation.

%%%%%%%%%%%%%%%%%%%%%%%%%%%%%%%%%%%%%%%%%%%%%%%%%%%%%%%%%%
\subsection{Standard nonlinear step}
\label{sec:sns}
%
%
%
%%%%%%%%%%%%%%%%%%%%%%%%%%%%%%%%%%%%%%%%%%%%%%%%%%%%%%%%%%

During the standard nonlinear step, we solve Eqns (\ref{eq:mhd1_rho}) through (\ref{eq:mhd1_B}) without the linear advection operator $(\vec{w}\cdot\nabla)$.
The resulting system is equivalent to the regular MHD equations written in terms of the residual velocity $\vec{v}' = \vec{v}-\vec{w}$ \emph{plus} the additional source terms $\vec{S}'_m$ and $S'_E$ given, respectively, by Eq. (\ref{eq:Sm0}) and Eq. (\ref{eq:SE0}).

This entitles us to employ the formalism already developed for the solution of the MHD equations using any stable conservative finite-volume or finite-difference discretization available with the PLUTO code.
Accordingly, we update each conserved quantity $q$ using the general building block
\begin{equation}\label{eq:build_block}
  \frac{q^{n+1} - q^n}{\Delta t^n} = \Big[-\nabla\cdot\vec{F}_q + S_q\Big]^n\,,
\end{equation}
which is the discrete version of Eq. (\ref{eq:SNS}).
In Eq. (\ref{eq:build_block}), the flux $\vec{F}_q$ follows from the solution of Riemann problems at zone interfaces, while the divergence term is expressed as the sum of two-point difference operators in each direction $d$.
Dropping the index $q$ for simplicity and considering a generic flux $\vec{F}$ with components $F_d=\vec{F}\cdot\hvec{e}_d$, we use
\begin{equation}
  \Big[\nabla\cdot\vec{F}\Big]^n = \sum_d\frac{1}{\Delta{\cal V}_d}\left(
  A_{d,+}F_{d,+} - A_{d,-}F_{d,-}\right)\,,
\end{equation}
where $F_{d,\pm}$ is the flux through the upper (+) and lower (-) cell boundaries with surface normal $\hvec{e}_d$, while $A_d$ and ${\cal V}_d$ are the interface areas and cell volume.

The magnetic field is evolved using the constrained transport formalism \citep[see, e.g., the papers by][]{PLUTO.2007,Flock_etal.2010}.

%%%%%%%%%%%%%%%%%%%%%%%%%%%%%%%%%%%%%%%%%%%%%%%%%%%%%%%%%%
\subsubsection{Conservative formulation}
\label{sec:source_terms}
%
%
%
%%%%%%%%%%%%%%%%%%%%%%%%%%%%%%%%%%%%%%%%%%%%%%%%%%%%%%%%%%

As anticipated in \S\ref{sec:FARGO}, the source term $S_q$ contains both point-wise contributions (e.g. gravitational or curvilinear terms) and terms involving the derivatives of the orbital velocity $\vec{w}$.  
Point-wise source terms not involving spatial derivatives are added to the right-hand side of the equations in a standard explicit way.
Conversely, source terms contributing to both the azimuthal momentum component and the energy equation require some additional considerations.
To this end, we consider the net change in the total azimuthal momentum during a single time-step update.
Using, for simplicity, Cartesian coordinates and neglecting gravity, a straightforward combination of the equations of continuity and the $y-$component of momentum yields
\begin{equation}\label{eq:tot_AM}
 (\rho v_y)^{n+1} = (\rho v_y)^n - \Delta t 
     \Big[\nabla\cdot\vec{F}_{m'_y} + w\nabla\cdot\vec{F}_\rho 
    + \rho\vec{v}\cdot\nabla w\Big]^n\,.
\end{equation}
Although the sum of the last two terms on the right-hand side is equal to $\nabla\cdot(w\vec{F}_\rho)$ at the continuous level, this may not necessarily hold in a discrete sense.
At the numerical level one should indeed expect $|w\nabla\cdot\vec{F}_\rho + \rho\vec{v}'\cdot\nabla w - \nabla\cdot(w\vec{F}_\rho)| = O(\Delta x^s)$, where $O(\Delta x^s)$ is the truncation error of the scheme.
As a consequence, total (linear or angular) momentum and, similarly, total energy density will be conserved only at the truncation level of the scheme.

We, instead, seek a numerical discretization that allows to restore the exact conservation of total linear (for Cartesian geometry) or angular (for polar grids) momentum and energy, also at the numerical level.
For this purpose, we step back in the derivation from Eqns (\ref{eq:mhd_rho})-(\ref{eq:mhd_B}) to (\ref{eq:mhd1_rho})-(\ref{eq:mhd1_B}) and note that the source term, as given in Eq. (\ref{eq:Sm0}), of the momentum equation simply results from the algebraic manipulation of
\begin{equation}\label{eq:Sm1}
 \vec{S}'_{m} = - \nabla\cdot(\rho\vec{v}'\vec{w}) 
             + \vec{w}\nabla\cdot(\rho\vec{v}')
             - \rho\vec{w}\cdot\nabla\vec{w} \,,
\end{equation}
where the last term is identically zero in Cartesian coordinates and only contributes to the radial-momentum source term in polar and spherical coordinates.
Similarly, one can show, after some algebra, that the source term in Eq. (\ref{eq:SE0}) of the energy equation results from
\begin{equation}\label{eq:SE1}
\begin{array}{lcl}
 S'_E &=& \DS \vec{w}\cdot\Big[
  \nabla\cdot(\rho\vec{v}'\vec{v}' - \vec{B}\vec{B} + \rho\vec{v}'\vec{w})
  \Big] 
  - \frac{w^2}{2}\nabla\cdot(\rho\vec{v}') 
\\ \noalign{\medskip}
 & &\DS
  - \nabla\cdot\left[\frac{\rho w^2}{2}\vec{v}' + \vec{w}\cdot\left(
     \rho\vec{v}'\vec{v}' - \vec{B}\vec{B}\right)\right]\,.
\end{array}
\end{equation}
This suggests that Eq. (\ref{eq:Sm1}) and (\ref{eq:SE1}) may be conveniently expressed in terms of the density and $y$-momentum upwind fluxes already at disposal during the conservative update.
For instance, we update density, (residual) azimuthal-momentum, and (residual) energy as
\begin{equation}\label{eq:rho_update}
  \frac{\rho^{n+1} - \rho^n}{\Delta t^n} =
 \Big[ -\nabla\cdot\vec{F}_\rho\Big]^n
\end{equation}

\begin{equation}\label{eq:my_update}
  \frac{(m'_y)^{n+1} - (m'_y)^n}{\Delta t^n} = \Big[
- \nabla\cdot\left(\vec{F}_{m_y}+w\vec{F}_\rho\right) 
               + w\nabla\cdot\vec{F}_\rho\Big]^n
\end{equation}

\begin{equation}\label{eq:E_update}
\begin{array}{lcl}
\DS\frac{(E')^{n+1} - (E')^n}{\Delta t^n} &=&\DS\left[-\nabla\cdot\left(
  \vec{F}_{E} + w\vec{F}_{m_y} + \frac{w^2}{2}\vec{F}_{\rho}
\right) + \right. 
\\ \noalign{\medskip}
  &   &\DS\left. + w\nabla\cdot\left(\vec{F}_{m_y} + w\vec{F}_\rho\right)
        - \frac{w^2}{2}\nabla\cdot\vec{F}_\rho\right]^n
\end{array}
\end{equation}
which shows that, by adding the product of Eq. (\ref{eq:rho_update}) and $w$ to Eq. (\ref{eq:my_update}), one obtains the conservative update of the total momentum with corresponding flux $F_{m_y}+wF_\rho$.
Since similar arguments hold by combining the density and momentum equations with Eq. (\ref{eq:E_update}), the resulting discretization enforces the conservations of both total momentum and total energy in both a continuous \emph{and} a numerical sense. 
The corresponding expressions for polar and spherical coordinates are a straightforward extension of these equations leading to both total angular momentum and total energy conservation. 
We report them in \S\ref{sec:polcoords} and \S\ref{sec:sphcoords}, respectively.
%The employment of upwind fluxes in the computation of $\vec{S}'_m$ and $S'_E$ confer more robustness to the numerical scheme.
We point out that the importance of a conservative formulation of the equations has already been recognized by other authors, e.g. \cite{Kley.1998}, who has shown that the inclusion of non-intertial forces as a source term may lead to erroneous results.

%Although Equations (\ref{eq:Sm1}) and (\ref{eq:SE1}) are apparently more 
%elaborate than their simplified verions, Eq. (\ref{eq:Sm0}) and 
%(\ref{eq:SE1}), they can 
%%%%%%%%%%%%%%%%%%%%%%%%%%%%%%%%%%%%%%%%%%%%%%%%%%%%%%%%%%%%%%%%%%%%%%%%%
\subsection{Linear transport step}
\label{sec:lts}
%
%
%%%%%%%%%%%%%%%%%%%%%%%%%%%%%%%%%%%%%%%%%%%%%%%%%%%%%%%%%%%%%%%%%%%%%%%%%

As anticipated, the solution of the linear transport equation in Eq. (\ref{eq:LTS}) consists of a uniform shift $|\vec{w}\Delta t|$ of the conserved variable profiles in the direction of orbital motion.
If the flow is predominantly aligned with one of the coordinate axes, say $\hvec{y}$, we can write the solution of Eq. (\ref{eq:LTS}) for $t\in[t^n,t^{n+1}]$ as $q(y,t) = q(y-w(t-t^n),t^n)$.
Extensions to polar and spherical coordinates are straightforward, provided that $y\to\phi$ and $wt\to\Omega t$.

Integration of Eq. (\ref{eq:LTS}) for a time step $\Delta t$ and over a cell size $\Delta y$ gives
\begin{equation}\label{eq:transport_int}
  q^{n+1}_j = q^n_j - \frac{1}{\Delta y}\left(
    \int_{y_{j+\HALF}-w\Delta t}^{y_{j+\HALF}} q^n(\xi)\,d\xi
 -  \int_{y_{j-\HALF}-w\Delta t}^{y_{j-\HALF}} q^n(\xi)\,d\xi\right)\,,
\end{equation}
where $q^n(\xi)=q(\xi,t^n)$ and $w$ does not depend on $y$.
The integrals on the right-hand side of Eq. (\ref{eq:transport_int}) can be converted into a finite sum corresponding to an integer shift of cells $m = {\rm floor}(w\Delta t/\Delta y + 1/2)$ plus a fractional remainder $\delta y = w\Delta t - m\Delta y$.
The final result may be cast as a two-point flux difference scheme
\begin{equation}\label{eq:LTS_Solw}
 q^{n+1}_{j} = q^n_{j_m} - \frac{\delta y}{\Delta y}\left(
   H_{j_m+\HALF} - H_{j_m-\HALF}\right)\,,
\end{equation}
where $j_m=j-m$ and
\begin{equation}\label{eq:Hflux}
 H_{j+\HALF} = \frac{1}{\delta y}
               \int_{y_{j+\HALF}-\delta y}^{y_{j+\HALF}}q^n(\xi)d\xi\
\end{equation}
is the upwind numerical flux.
We remark that the finite summations corresponding to an integer cell shift (implicitly contained in the integral of Eq. \ref{eq:transport_int}) do not need to be explicitly computed since most terms cancel out when taking their difference. 
We also note that, since $|\delta y| < \Delta y$ by construction, the scheme is always stable regardless of the choice of $\Delta t$.

The fluxes given by Eq. (\ref{eq:Hflux}) may be computed to an arbitrary order of accuracy by assuming a piecewise polynomial representation of the data inside the cell.
If, for instance, we assume a piecewise linear distribution of $q$ inside the zone, then the integral in Eq. (\ref{eq:Hflux}) takes the form
\begin{equation}\label{eq:LTS_MH_flux}
 H_{j+\HALF} = 
\left\{\begin{array}{lcl}
 \DS q_j + \frac{\Delta q_j}{2}\left(1 - \frac{\delta y}{\Delta y}\right)
 & \quad\mathrm{if}\quad& \delta y \ge 0 \,,
\\ \noalign{\bigskip}
\DS q_{j+1} - \frac{\Delta q_{j+1}}{2}\left(1 + \frac{\delta y}{\Delta y}\right)
 & \quad\mathrm{if}\quad& \delta y \le 0 \,,
\end{array}\right.
\end{equation}
where $\Delta q_j$ is computed using a standard slope limiter.
Equation (\ref{eq:LTS_MH_flux}) corresponds to the well-known second-order MUSCL-Hancock scheme of \citet{vanLeer.1984}.
Similarly, using a piecewise parabolic distribution inside the cell we have
\begin{equation}\label{eq:LTS_PPM_flux}
 H_{j+\HALF} = 
\left\{\begin{array}{lcl}
\DS q_{j,+} - \frac{\delta y}{\Delta y}\left[
   \frac{\delta q_j}{2} 
  + \delta^2q_j\left(\frac{3}{2} - \frac{\delta y}{\Delta y}\right)\right] 
 & \;\mathrm{if}\;& \delta y \ge 0 \,,
\\ \noalign{\bigskip}
\DS q_{j+1,-} - \frac{\delta y}{\Delta y}\left[
   \frac{\delta q_{j+1}}{2} 
  - \delta^2q_{j+1}\left(\frac{3}{2} + \frac{\delta y}{\Delta y}
  \right)\right]
 & \; \mathrm{if}\;& \delta y \le 0 \,,
\end{array}\right.
\end{equation}
where $q_{j,\pm}$ are third (or higher) order rightmost and leftmost interpolated values with respect to the cell center, $\delta q_j = q_{j,+}-q_{j,-}$ and $\delta^2q_j = q_{j,+} - 2q_j + q_{j,-}$. 
This is essentially the PPM scheme for a linear advection equation \citep{ColWoo.1984} and is our default choice, unless stated otherwise.

%%%%%%%%%%%%%%%%%%%%%%%%%%%%%%%%%%%%%%%%%%%%%%%%%%%%%%%%%%%%%%%%%%%%%%%%%
\subsubsection{Magnetic field transport}
%
%
%%%%%%%%%%%%%%%%%%%%%%%%%%%%%%%%%%%%%%%%%%%%%%%%%%%%%%%%%%%%%%%%%%%%%%%%%

In the constrained transport formalism, the three components of magnetic field are discretized on a staggered mesh and evolved as surface averages placed at the different zone faces to which they are orthogonal.
In this sense, we locate the components of $\vec{B}$ by means of a half-integer subscript, that is, $B_{x,i+\HALF,j,k}$, $B_{y,i,j+\HALF,k}$ and $B_{z,i,j,k+\HALF}$ to denote the magnetic field components centered on the $x$, $y$, and $z$ faces, respectively.

The evolution is carried out using a discrete version of Stokes' theorem applied to Eq. (\ref{eq:mhd1_B}), where a line integral of the electric field is properly evaluated at zone edges. 
This guarantees that the divergence-free condition of the magnetic field is also maintained to machine precision during the linear transport step.
During the transport step (i.e. when $\vec{v}'=\vec{0}$), this leads to the following discretization of Eq. (\ref{eq:mhd1_B})
\begin{equation}\label{eq:Bx_update}
  B_{x,i+\HALF}^{n+1} = B^n_{x,i+\HALF} - \frac{
  {\cal E}^z_{i+\HALF,j+\HALF}-{\cal E}^z_{i+\HALF,j-\HALF}}{\Delta y_j}
\end{equation}
\begin{equation}\label{eq:By_update}
  B^{n+1}_{y,j+\HALF} = B^n_{y,j+\HALF} 
- \frac{{\cal E}^x_{j+\HALF,k+\HALF}-{\cal E}^x_{j+\HALF,k-\HALF}}{\Delta z_k}
+ \frac{{\cal E}^z_{i+\HALF,j+\HALF}-{\cal E}^z_{i-\HALF,j+\HALF}}{\Delta x_i}
\end{equation}
\begin{equation}\label{eq:Bz_update}
  B^{n+1}_{z,k+\HALF} = B^n_{z,k+\HALF} + \frac{
  {\cal E}^x_{j+\HALF,k+\HALF}-{\cal E}^x_{j-\HALF,k+\HALF}}{\Delta y_j}\,,
\end{equation}
where ${\cal E}^x=-\int wB_z\,dt$ and ${\cal E}^z=\int wB_x\,dt$ are the $x$ and $z$ components of the time-integrated electromotive force computed, in analogy with the results of the previous section, as
\begin{equation}\label{eq:Ez}
  \E^z_{i+\HALF,j+\HALF} =\delta yH_{j_m+\HALF} + \left\{\begin{array}{cl}
\DS  \Delta y\sum_{j'=j_m+1}^{j} B^n_{x,i+\HALF,j'} & \;\mathrm{if}\quad m > 0 
\\ \noalign{\medskip}
 \DS 0                                              &\;\mathrm{if}\quad m = 0 
\\ \noalign{\medskip}
\DS  -\Delta y\sum_{j'=j+1}^{j_m} B^n_{x,i+\HALF,j'}& \;{\rm if}\quad m < 0 
\end{array}\right.
\end{equation}
with $H_{j_m+\HALF}$ computed from Eq. (\ref{eq:Hflux}) using $q=B_x$ and
\begin{equation} \label{eq:Ex}
  \E^x_{j+\HALF,k+\HALF} = \delta yH_{j_m+\HALF} + \left\{\begin{array}{cl}
\DS - \Delta y\sum_{j'=j_m+1}^{j} B^n_{z,j',k+\HALF}&\;\mathrm{if}\quad m > 0
\\ \noalign{\medskip}
\DS  0                                              &\;\mathrm{if}\quad m = 0
\\ \noalign{\medskip}
\DS   \Delta y\sum_{j'=j+1}^{j_m} B^n_{z,j',k+\HALF}& \;\mathrm{for}\quad m < 0
\end{array}\right.
\end{equation}
with $H_{j_m+\HALF}$ computed from Eq. (\ref{eq:Hflux}) with $q=-B_z$.
In the previous equations, we have dropped, for simplicity, the full subscript notations when redundant and kept only the half-increment notation to represent the different magnetic and electric field components.

We point out that the updating formulas for $B_x$ and $B_z$ could be implemented in exactly the same way as done for Eq. (\ref{eq:LTS_Solw}) since the differences of fluxes along the y-direction, defined by Eq. (\ref{eq:Bx_update}) and (\ref{eq:Bz_update}), leads to the cancellation of most terms leaving only the upstream values.
On the other hand, cancellation does not occur when updating $B_y$, since Eq. (\ref{eq:By_update}) involves differences of terms along the $x$ and $z$ direction corresponding to the winding of the field under the action of the shear. 
For this reason, the full summation must now be retained in Eq. (\ref{eq:Ez}) and (\ref{eq:Ex}) when evaluating the time-integrated electromotive force.

For the sake of completeness, we also report the corresponding update expressions for other geometries. 
In polar coordinates $(R,\phi, z)$, we have
\begin{equation}\label{eq:Bx_update_pol}
B^{n+1}_{R,i+\HALF} = B^n_{R,i+\HALF} 
 - \frac{\E^z_{i+\HALF,j+\HALF}-\E^z_{i+\HALF,j-\HALF}}{\Delta\phi_j}
\end{equation}
\begin{equation}\label{eq:By_update_polar}\begin{array}{ll}
B^{n+1}_{\phi,j+\HALF} =  B^n_{\phi,j+\HALF} 
   &\DS -\, R_i\frac{\E^R_{j+\HALF,k+\HALF}-\E^R_{j+\HALF,k-\HALF}}
             {\Delta z_k}
\\ \noalign{\medskip}
   &\DS +\, \frac{ R_{i+\HALF}\E^z_{i+\HALF,j+\HALF}
                -R_{i-\HALF}\E^z_{i-\HALF,j+\HALF}}{\Delta R_i} 
\end{array}
\end{equation}
\begin{equation}\label{eq:Bz_update_polar}
 B^{n+1}_{z,k+\HALF} = B^n_{z,k+\HALF} 
   + \frac{\E^R_{j+\HALF,k+\HALF} - \E^R_{j-\HALF,k+\HALF}}{\Delta\phi_j}\,,
\end{equation}
where ${\cal E}^R = -\int \Omega B_z\,dt$, ${\cal E}^z =  \int \Omega B_R\,dt$ (with $\Omega = w/R$) are computed similarly to Eq. (\ref{eq:Ez}) and (\ref{eq:Ex}).% with $\Delta y \to \Delta \phi$, $B_x\to B_R$, $\E^x\to\E^R$.

Likewise, in spherical coordinates $(r,\theta,\phi)$ we have
\begin{equation}\label{eq:Bx_update_sph}
B^{n+1}_{r,i+\HALF} = B^n_{r,i+\HALF} 
   + \frac{ {\cal E}^\theta_{i+\HALF,k+\HALF}
           -{\cal E}^\theta_{i+\HALF,k-\HALF}}{\Delta\phi_k}
\end{equation}
\begin{equation}\label{eq:By_update_sph}
B^{n+1}_{\theta,j+\HALF} = B^n_{\theta,j+\HALF} 
   - \frac{ {\cal E}^r_{j+\HALF,k+\HALF}
   - {\cal E}^r_{j+\HALF,k-\HALF}}{\Delta\phi_k}
\end{equation}
\begin{equation}\label{eq:Bz_update_sph}\begin{array}{ll}
B^{n+1}_{\phi,k+\HALF} = B^n_{\phi,k+\HALF} 
   & \DS -\, \frac{\sin\theta_{j+\HALF}{\cal E}^r_{j+\HALF,k+\HALF}
                -\sin\theta_{j-\HALF}{\cal E}^r_{j-\HALF,k+\HALF}}{\Delta\theta_j}
\\ \noalign{\medskip}
   & \DS +\, \sin\theta_j
  \frac{ r^2_{i+\HALF}{\cal E}^\theta_{i+\HALF,k+\HALF}
       - r^2_{i-\HALF}{\cal E}^\theta_{i-\HALF,k+\HALF}}{r_i\Delta r_i}\,,
\end{array}
\end{equation}
where $\E^r =  \int \Omega B_\theta\,dt$, ${\cal E}^\theta = -\int \Omega B_r\,dt$, and $\Omega = w/(r\sin\theta)$ are computed similarly to Eqs. (\ref{eq:Ez}) and (\ref{eq:Ex}).% with $\Delta y\to \Delta\phi$,$B_x\to B_r$.

%%%%%%%%%%%%%%%%%%%%%%%%%%%%%%%%%%%%%%%%%%%%%%%%%%%%%%%%%%%%%%
\subsection{Expected speedup}
%
%
%%%%%%%%%%%%%%%%%%%%%%%%%%%%%%%%%%%%%%%%%%%%%%%%%%%%%%%%%%%%%%

In general, the expected time-step gain is problem-dependent but can also be affected by several factors such as the magnitude of the orbital velocity, the grid geometry, and the cell aspect ratio.
An estimate of the speedup may be directly computed from Eq. (\ref{eq:Ca}) by taking the ratio of the time step obtained with orbital advection to that obtained without.
Specializing, for example, to polar coordinates and RK2 time-stepping, one can estimate a time-step increase
\begin{equation}
  \frac{\Delta t_F}{\Delta t_s} = 
\frac{\DS \max_{ijk}
 \left[  \frac{|v_R|+c_{f,R}}{\Delta R} +
        +\frac{|v'_\phi + w| + c_{f,\phi}}{R\Delta\phi}
	+\frac{|v_z| + c_{f,z}}{\Delta z}\right]}
{\DS \max_{ijk}\left[\frac{|v_R| + c_{f,R}}{\Delta R} 
                    +\frac{|v'_\phi|+c_{f,\phi}}{R\Delta\phi} 
                    +\frac{|v_z| + c_{f,z}}{\Delta z}\right]}
\,.
\end{equation}
If the characteristic signal fluctuations are of comparable magnitude, $|v_r| + c_{f,r} \approx |v'_\phi| + c_{f,\phi} \approx |v_z| + c_{f,z} \approx \lambda'$ and are such that $\lambda'\ll |w|$, the previous expression simplifies to
\begin{equation}
\frac{\Delta t_F}{\Delta t_s} \approx
 \frac{\DS \max_{ijk}
  \left[\frac{1}{\Delta R} 
      + \frac{M+1}{R\Delta\phi} 
      + \frac{1}{\Delta z}\right]}
{\DS \max_{ijk}\left[\frac{1}{\Delta R} 
                   + \frac{1}{R\Delta\phi} 
                   + \frac{1}{\Delta z}\right]}\,,
\end{equation}
where $M = |w|/|\lambda'|$ should be of the same order as the fast magnetosonic Mach number.
The previous estimate shows that a sensible choice of the grid resolution has a direct impact on the expected gain: a finer resolution in the azimuthal direction ($R\Delta\phi\ll \Delta R$) favors a larger gain, whereas cells with a smaller radial extent may become less advantageous.

%%%%%%%%%%%%%%%%%%%%%%%%%%%%%%%%%%%%%%%%%%%%%%%%%%%%%%%%%%%%%%%%%
\subsection{Parallel implementation}
%
%
%
%%%%%%%%%%%%%%%%%%%%%%%%%%%%%%%%%%%%%%%%%%%%%%%%%%%%%%%%%%%%%%%%%

In parallel computations, the simulation domain is decomposed into smaller sub-grids, which are then solved simultaneously by several processing units.
For time-explicit calculations, only boundary data stored in the ghost zones have to be exchanged between neighboring processors.
With FARGO-MHD, however, this approach presents some difficulties since the linear transport step may involve shifts of an arbitrary number of zones along the orbital direction. 
This implies that data values could in principle be exchanged between all processors lying on the same row of a parallel domain decomposition, thus involving many more communications than a regular exchange of ghost zones between adjacent processors.
%Although one could avoid parallelization of the orbital direction so that each sub-domain contains the full azimuthal range, this strategy is not efficient on architectures with several thousands of processors.
%On the other side, intensive all-to-all parallel communications that may be %required by an arbitrary displacement of zones in one direction can %severely degrade the computational performance.

In the tests and applications presented here, we observed that the maximum zone shift $m_{\max}$ hardly ever exceeds the typical grid size $N_\phi$ of a single processor. 
As efficient parallel applications require $N_\phi \gtrsim 8-16$, we can safely assume that the condition $m_{\max} \le N_\phi$ is virtually always respected and does not represent a stringent requirement for most applications.
In such a way, parallel communications are performed as a cyclic shift between neighboring processes only along the orbital direction and the computational overhead becomes approximately $1+m_{\max}/n_g$ times larger than a regular boundary call (where $n_g$ is the number of ghost zones) in any direction.
Notice also that, when the orbital velocity does not change sign across the domain, the amount of subtask communication can be halved since information always travels in the same direction and data values need to be transferred from one processor to the next following the same pattern.

%%%%%%%%%%%%%%%%%%%%%%%%%%%%%%%%%%%%%%%%%%%%%%%%%%%%%%%%%%%%%%%%%%%%%%%%%
\section{Numerical benchmarks}
\label{sec:tests}
%
%
% Cartesian Runs: use PPM+Char_Tracing (new version), 
%                 with PARABOLIC_LIM 1, REF_STATE     3
%
%
%
%%%%%%%%%%%%%%%%%%%%%%%%%%%%%%%%%%%%%%%%%%%%%%%%%%%%%%%%%%%%%%%%%%%%%%%%%

We present a number of hydrodynamical and magnetohydrodynamical test problems where the proposed orbital advection algorithm is directly compared with the standard traditional integration method.
Unless stated differently, we make use of the ideal EoS with $\Gamma=5/3$ and the PPM method, given by Eq. (\ref{eq:LTS_PPM_flux}), is the default interpolation during the linear transport step.

%%%%%%%%%%%%%%%%%%%%%%%%%%%%%%%%%%%%%%%%%%%%%%%%%%%%%%%%%%%%%%%%%
\subsection{Vortex dynamics in a two-dimensional Keplerian flow}
%
%
%
%%%%%%%%%%%%%%%%%%%%%%%%%%%%%%%%%%%%%%%%%%%%%%%%%%%%%%%%%%%%%%%%%

\begin{figure}\centering
\includegraphics[width=0.45\textwidth]{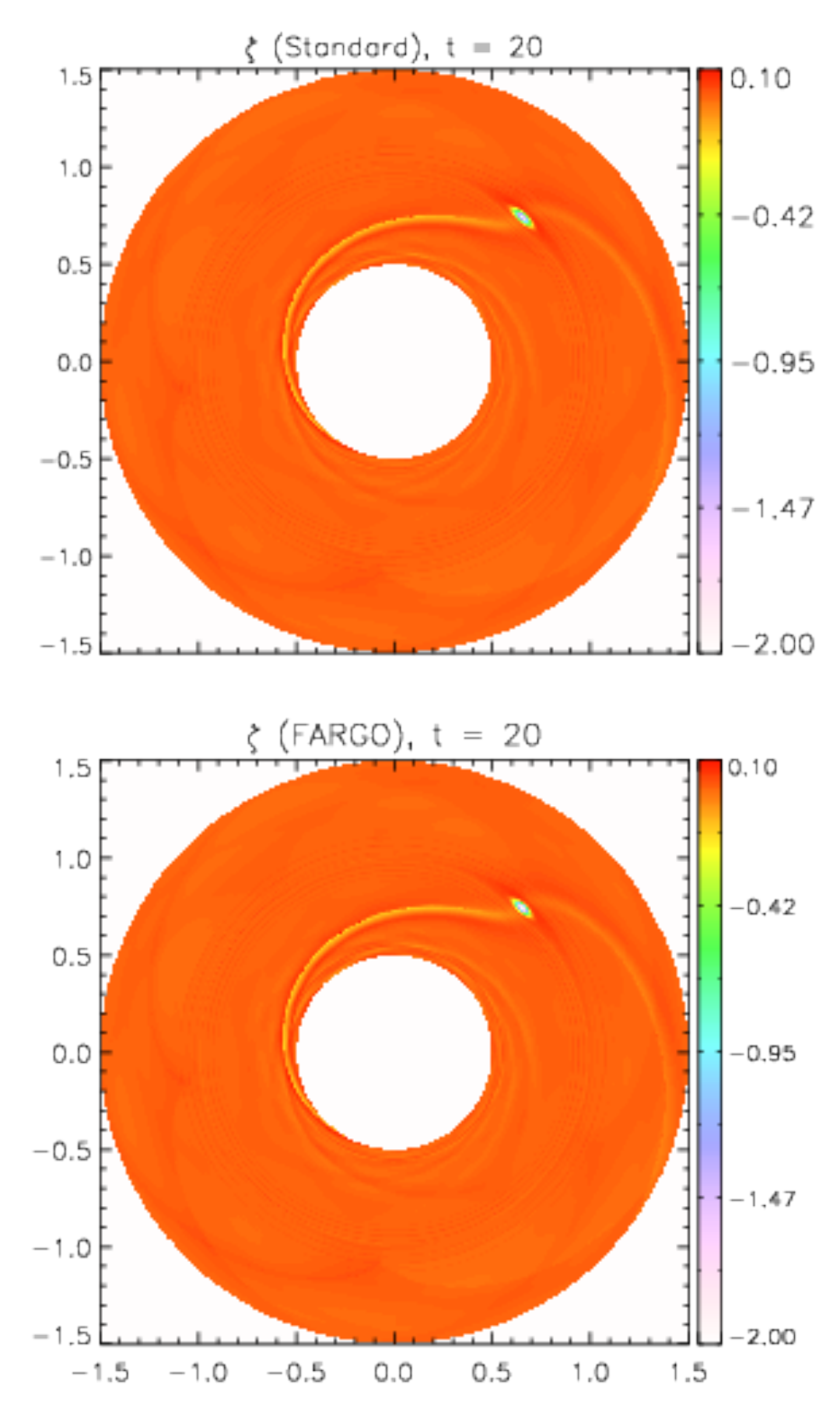}
\caption{\footnotesize Local potential vorticity for the hydrodynamical vortex test after 20 orbits at the resolution of $1024\times 4096$. Top panel: standard integration. Bottom panel: orbital advection.}         
\label{fig:Disk2D-pvort}
\end{figure}
\begin{figure}[!ht]\centering
\includegraphics[width=0.45\textwidth]{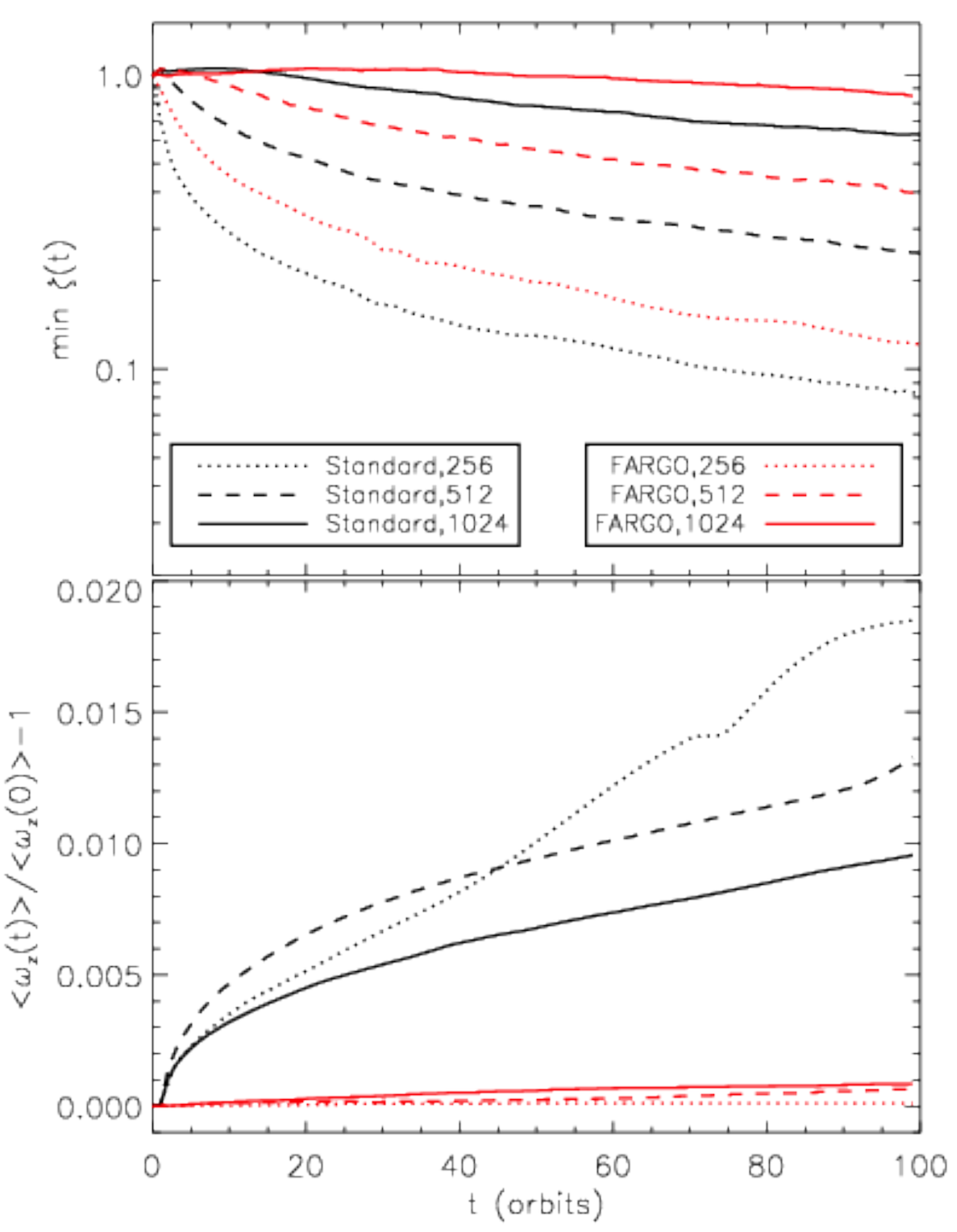}
\caption{\footnotesize Time evolution of the vortensity minimum (top, normalized to its initial value) and relative variation in the total integrated vorticity $\av{\omega_z(t)}=\av{\hvec{z}\cdot\nabla\times\vec{v}(t)}$ (bottom) for the hydrodynamical vortex problem. 
Dotted, dashed, and solid lines refer to computations obtained at the resolutions of $256\times 1024$, $512\times 2048$, and $1024\times 4096$ grid zones, respectively. Black and red colors correspond to the the standard integration algorithm and FARGO-MHD, respectively.}        
\label{fig:Disk2D-hist}
\end{figure}

We begin by considering the dynamical evolution of an anti-cyclonic vortex in a two-dimensional (2D) Keplerian disk using polar coordinates $(R,\phi)$.
The initial background state is defined by constant density and pressure equal respectively to $\rho=1$ and $p=1/(\Gamma M^2)$, where $M=10$ is the Mach number at $R=1$. 
The disk rotates with angular velocity $\Omega(R) =R^{-3/2}$ under the influence of a point-mass gravity $\Phi = -1/R$ and fills the computational domain defined by $0.4\le R\le 2$ and $0\le\phi\le 2\pi$. 
With these definitions, the disk scale-height is given by $H(R)=c_s/\Omega(R)$, where $c_s = 1/M$.
A circular vortex is superposed on the mean Keplerian flow and initially defined by
\begin{equation}
 \left(\begin{array}{c}
 \delta v_R    \\ \noalign{\medskip}
 \delta v_\phi 
 \end{array}\right)
  = 
\kappa\exp\left(-\frac{x^2 + y^2}{h^2}\right)
 \left(\begin{array}{cc}
 \cos\phi  & \sin\phi \\ \noalign{\medskip}
 -\sin\phi & \cos\phi
 \end{array}\right)
\left(\begin{array}{c}
 -y \\ \noalign{\medskip}
  x
\end{array}\right) \,,
\end{equation}
where $h=H(1)/2$ is the size of the vortex in terms of the local scale-height at $R=1$, $\kappa=-1$ is the vortex amplitude, and $x=R\cos\phi-R_0\cos\phi_0$ and $y=R\sin\phi-R_0\sin\phi_0$ are the Cartesian coordinates measured from the center of the vortex initially located at $R_0=1$ and $\phi_0=\pi/4$.

We solve the hydrodynamical equations (no magnetic field) using the second-order Runga-Kutta time-stepping with $C_a=0.4$ using linear reconstruction and follow the system until $t=200\pi$, i.e. for 100 revolutions of the ring at $R=1$. 
Computations are carried out with and without orbital advection at three different resolutions corresponding to $256\times 1024$ (low), $512\times 2048$ (medium), and $1024\times 4096$ (high), yielding approximately square cells in the proximity of the vortex.
To enforce conservation, we ensure that no net flux is established across the domain and therefore choose reflective boundary conditions at the inner and outer radial boundaries and impose periodicity along the $\phi$ direction.

\begin{table}[!h]
\caption{\footnotesize Average time step for the hydrodynamical vortex test problem at different resolutions using the standard scheme ($2^{\rm nd}$ column) and with orbital advection ($3^{\rm rd}$ column). The time step gain is reported in the last column.}
\label{tab:Disk2D-dt}
\centering
\begin{tabular}{cccc}\hline
$N_R\times N_\phi$    & $\Delta t$ (Standard) & $\Delta t$ (FARGO-MHD) & Gain 
\\ \hline\hline\noalign{\smallskip}
  $256\times1024$  & $1.15\,10^{-3}$ & $1.40\,10^{-2}$ & $12.07$\\ 
 \noalign{\smallskip}
  $512\times2048$  & $5.73\,10^{-4}$ & $6.61\,10^{-3}$ & $11.53$\\ \noalign{\smallskip}
  $1024\times4096$ & $2.84\,10^{-4}$ & $3.13\,10^{-3}$ & $10.99$\\ 
\hline
\end{tabular}
\end{table}

After a few revolutions, the vortex experiences a nonlinear adjustment to its final persistent structure. 
This process is accompanied by the emission of spiral density waves that radiate away the energy excess of the initial state \citep{BodTev_etal.2007}.
Figure \ref{fig:Disk2D-pvort} shows the local potential vorticity, or vortensity, defined as $\zeta = [\nabla\times(\vec{v}-\Omega R\hvec{\phi})]_z/\rho$ after $20$ revolutions using the standard integration algorithm (top) and the proposed FARGO scheme (bottom) at the highest resolution.
The two snapshots are in good agreement, although the simulation using orbital advection required only $\sim37,000$ steps compared to the $\sim444,000$ steps required by the standard computation. 
We found indeed that the time step increased, on average, by a factor between $\sim 11$ and $\sim 12$ with orbital advection, as reported in Table \ref{tab:Disk2D-dt}.

As an indicative measure of the dissipation properties of the scheme, we compare, in the top panel of Fig \ref{fig:Disk2D-hist}, the time history of the local vortensity minimum for the selected grid sizes.
An increase in the grid resolution favors a slower vortex decay leading to the formation of stable, long-lived vortex structures \citep{BodTev_etal.2007}. The same effect is obtained, at an equivalent resolution, by employing orbital advection.

Similarly, we inspected the relative variation in the total integrated vorticity presented as a function of time in the bottom panel of Fig \ref{fig:Disk2D-hist}.
Strictly speaking, we note that vorticity in a 2D compressible flow is a conserved quantity only if the fluid is barotropic, i.e., if $p=p(\rho)$.
Nevertheless, our results indicate that the amount of generated vorticity   decreases from $\approx 18\%$ (at the lowest resolution) to $\approx 10\%$ (at the largest) using the standard integration scheme (black lines), while it remains smaller than $1\%$ when orbital advection is employed (red lines).

Lastly, we verified that both total energy (including the gravitational contribution) and angular momentum are conserved to machine accuracy, as expected from the conservative formulation of our scheme.

%%%%%%%%%%%%%%%%%%%%%%%%%%%%%%%%%%%%%%%%%%%%%%%%%%%%%%%%%%%%
\subsection{Magnetohydrodynamical vortex in a shear flow}
%
%
%
%%%%%%%%%%%%%%%%%%%%%%%%%%%%%%%%%%%%%%%%%%%%%%%%%%%%%%%%%%%%

\begin{figure*}[ht!]\centering
\includegraphics[width=0.95\textwidth]{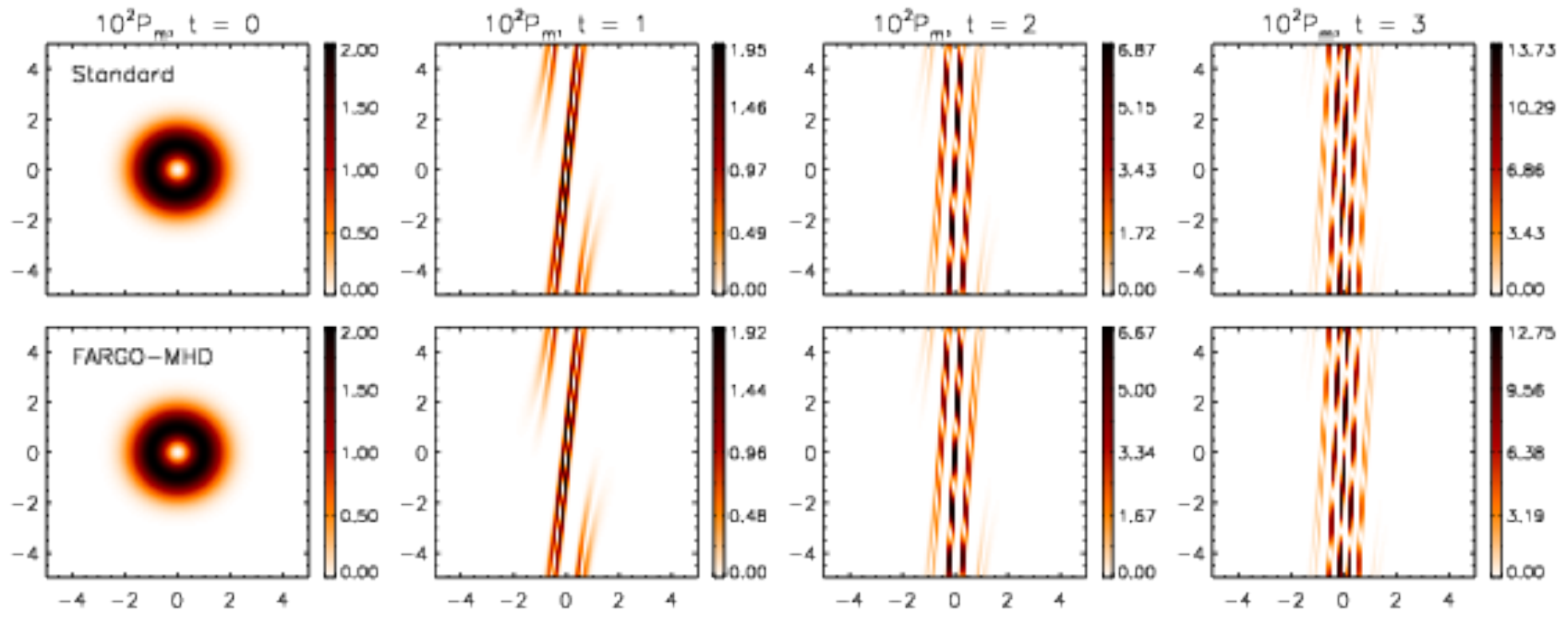}
\caption{\footnotesize Evolution of the magnetic pressure distribution, $P_m=(B_x^2+B_y^2)/2$ ($\times 100$), for the MHD vortex test computed using the standard MHD update (top row) and with FARGO-MHD (bottom row) using $512^2$ zones. From left to right, we show the evolution from time $t=0$ to $t=3$.}  
\label{fig:vortex-map}%
\end{figure*}

In the next example, we investigate the stretching of a 2D MHD vortex in a super-fast sheared flow using Cartesian coordinates.
The initial condition is similar to the one used by \cite{MigTzeBod.2010} and consists of a uniform-density ($\rho=1$), background shear flow with velocity profile $v_y = M\tanh(x)/2$, where $M$ is the sonic Mach number.
The magnetic field and pressure distributions are, respectively, given by
\begin{equation}\label{eq:vortex_B_and_p}
  (B_x,B_y) = (-y,x)\mu e^{(1-r^2)/2}\,,\quad
  p = \frac{1}{\Gamma} + \frac{1}{2}e^{1-r^2}\mu^2(1-r^2)\,,
\end{equation}
where $r^2=x^2+y^2$ and, for the present example, we adopt $\mu=0.02$ and $\Gamma=5/3$.

We choose the square region $x,y\in[-5,5]$ as our computational domain and impose periodic boundary conditions in the $y$-direction while applying reflecting conditions in the $x$-directions.
These choices ensure that no net flux is established across the domain boundaries so that density, momentum and energy are globally conserved.
We note that, in the absence of shear (for $M=0$), the previous configuration is an exact solution of the ideal MHD equations, representing a circular magnetic field-loop supported against a pressure gradient.
In the present context, however, we choose $M=20$ and follow the stretching of the initial loop configuration as time advances.
Computations are carried on $512^2$ grid zones using the PPM scheme with Courant number $C_a=0.8$.

\begin{figure}[!h]\centering
\includegraphics[width=0.5\textwidth]{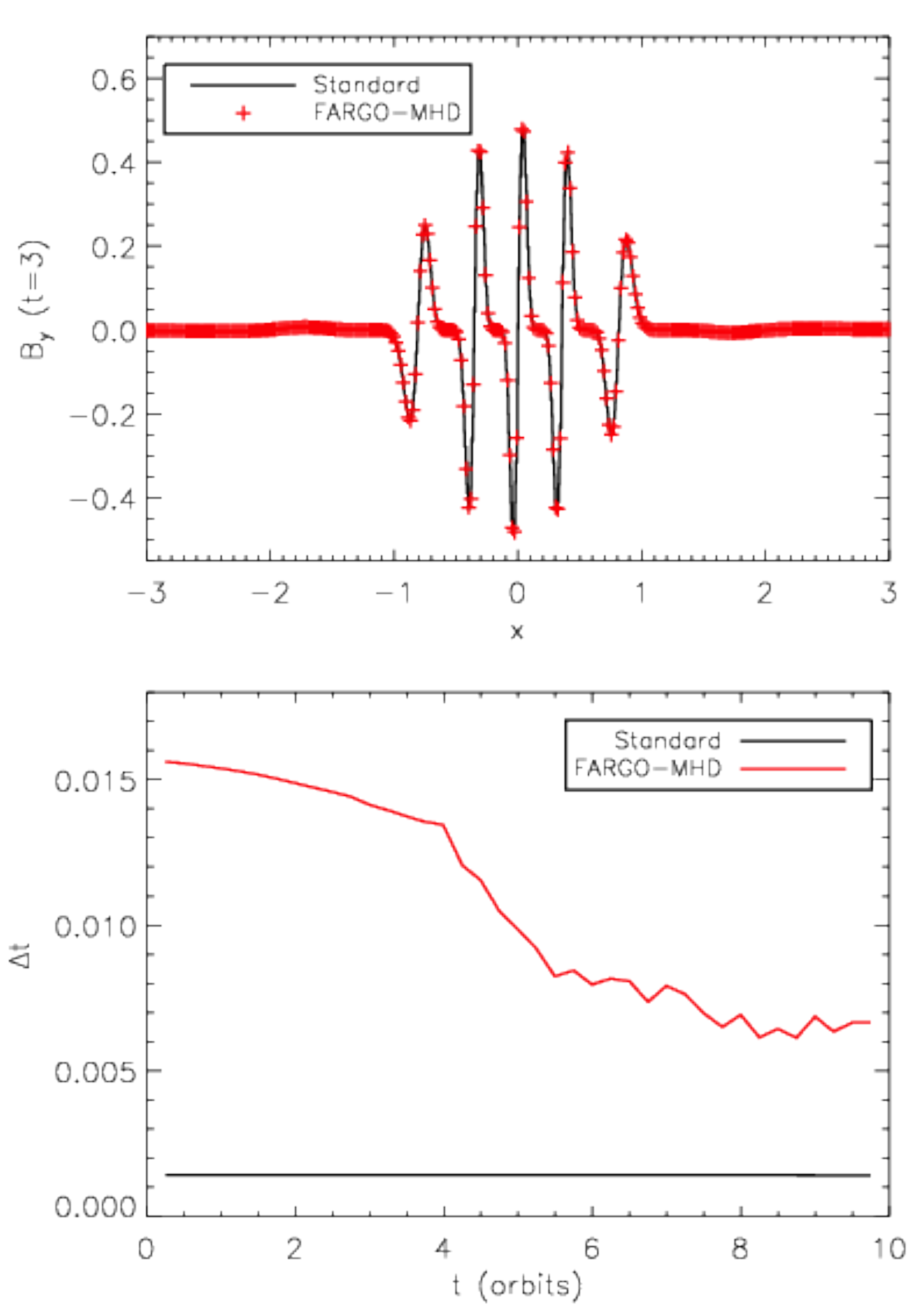}
\caption{\footnotesize Top panel: horizontal cuts at $y=0$ of the $y$ component of magnetic field at $t=3$ for the MHD vortex test in Cartesian coordinates using the standard integration method (solid line) and FARGO-MHD (plus signs). 
Bottom panel: time-step variation during the computation. Solid and dashed lines correspond to standard and FARGO-MHD computations, respectively.} 
\label{fig:vortex-by}
\end{figure}

Results with and without the FARGO-MHD algorithm are shown in Fig. \ref{fig:vortex-map}, where we compare the magnetic pressure distribution for $t=0,1,2,3$.
The two solutions look indistinguishable, as also visible from a one-dimensional cut of the y-component of magnetic field at $t=3$, shown in the left panel of Fig \ref{fig:vortex-by}.
A plot of the time step is shown in the right panel of the same figure, where one can see that orbital advection attains a larger gain ($\gtrsim 9$) during the first four revolutions and decreases to $\approx 4.5$ towards the end.
In terms of CPU time, the standard integration took approximately 2 hours and 31 minutes whereas only $26$ minutes were required using the orbital advection scheme.
This gives, on average, a speedup of $\approx 5.8$.

\begin{figure}[!h]\centering
\includegraphics[width=0.5\textwidth]{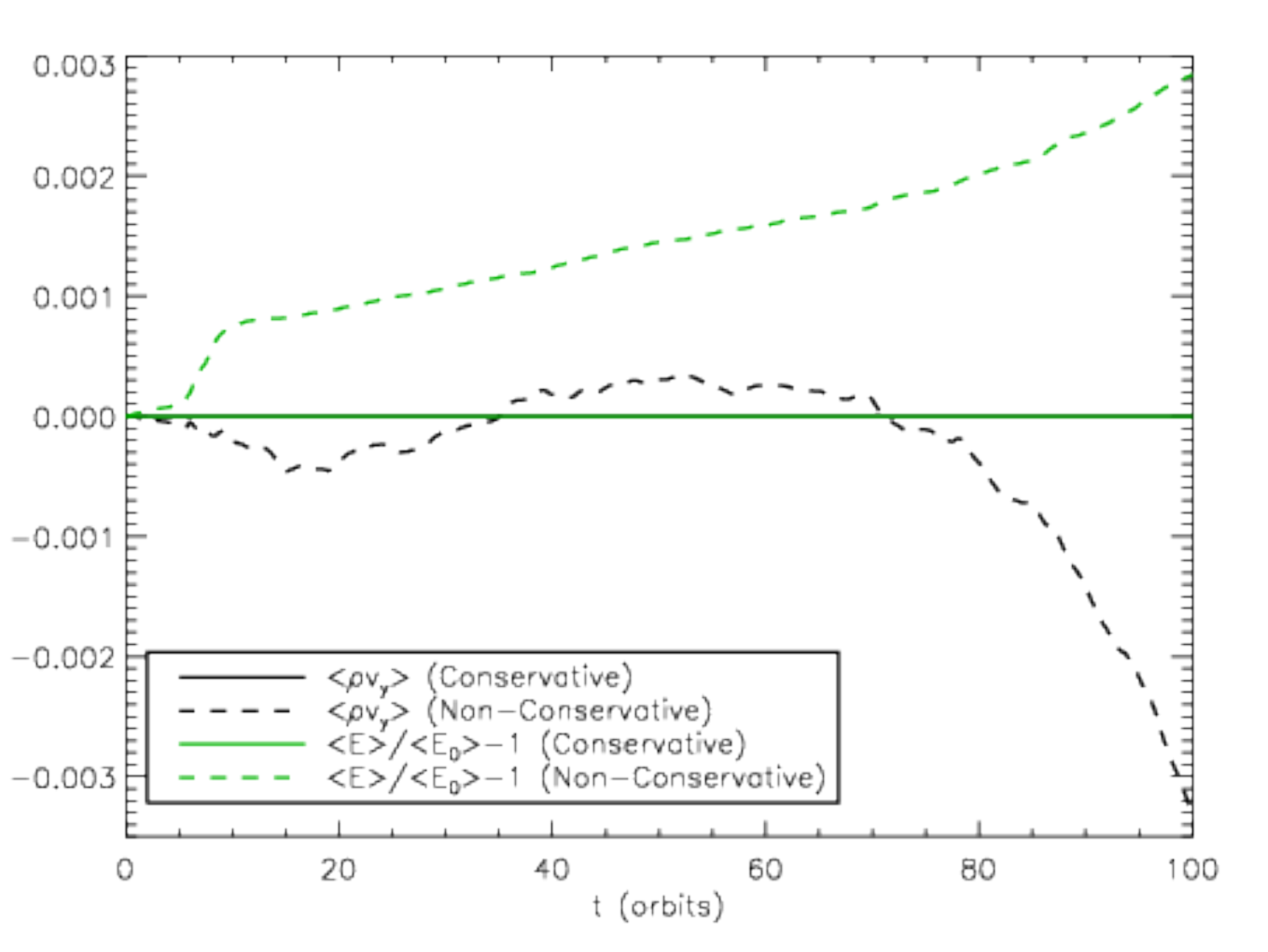}
\caption{\footnotesize Temporal evolution of the volume-averaged $y$ component of momentum (black) and fractional average energy variation (green) for the MHD vortex with random perturbation using the non-conservative (dashed lines) and conservative (solid line) versions of the orbital advection scheme.}         
\label{fig:vortex-rnd}
\end{figure}

To check whether total energy and angular momentum are conserved, we repeated the integration for a much longer time, corresponding to $100$ revolutions, and added a random horizontal velocity component of the form
\begin{equation}
  v_x = \left({\cal R}-\HALF\right)e^{-(x/2)^2}\,,
\end{equation}
where ${\cal R}$ is a random number between $0$ and $1$.
We compare, in Fig. \ref{fig:vortex-rnd}, the conservative and the non-conservative variants of the orbital advection algorithm by plotting the volume-average of the $y$ component of momentum and the relative variation in the total energy as time advances.
As expected, the former conserves momentum and energy to machine accuracy, while the latter exhibits increasing deviations from zero as the computation proceeds.

%

%%%%%%%%%%%%%%%%%%%%%%%%%%%%%%%%%%%%%%%%%%%%%%%%%%%%%%%%
\subsection{Shearing-box models}
\label{sec:shearingbox}
%
%
%
%%%%%%%%%%%%%%%%%%%%%%%%%%%%%%%%%%%%%%%%%%%%%%%%%%%%%%%%

The shearing-box approximation \citep{HawGamBal.1995} provides a local Cartesian model of a differentially rotating disk.
By neglecting curvature terms, one focuses on the evolution of a small rectangular portion of the disk in a frame of reference co-rotating with the disk at some fiducial radius where the orbital frequency is $\Omega_0$. 
In this approximation, the orbital motion is described by a linear velocity shear of the form $\vec{w}=-q\Omega_0x\hvec{y}$, where 
\begin{equation}\label{eq:SB_q}
 q = -\frac{1}{2}\frac{d\log\Omega^2(r)}{d\log r}
\end{equation}
is a local measure of the differential rotation.

The large-scale motion of the disk is described by assuming that identical boxes slide relative to the computational domain in the radial direction, a requisite implemented by the shearing-box boundary condition, which enforce sheared periodicity. 
For any flow quantity $q$ not containing $v_y$, this is formally expressed by
\begin{equation}\label{eq:SB_bc}
\begin{array}{lcl}
 q(x,y,z)   &\quad\to\quad& q(x+L_x,y-q\Omega_0 L_xt,z)
 \\ \noalign{\medskip}
 v_y(x,y,z) &\quad\to\quad& v_y(x+L_x,y-q\Omega_0 L_xt,z) + q\Omega_0 L_x\,,
\end{array}
\end{equation}
where $L_x$ is the radial ($x$) extent of the domain.
The implementation of the shearing-box boundary conditions is done similarly to \cite{GreZie.2007} and employ the same techniques described in this paper.

We report, for the sake of completeness, the basic equations in Appendix \ref{sec:SB} and refer the reader to \citet{HawGamBal.1995}, \citet{GreZie.2007} and \citet{StoGar.2010} for a thorough discussion.
In the tests below, we employ an isothermal EoS, assume a Keplerian flow ($q=3/2$), and adopt the unsplit CTU+PPM scheme with a Courant number $C_a = 0.45$.

%%%%%%%%%%%%%%%%%%%%%%%%%%%%%%%%%%%%%%%%%%%%%%%%%%%%%%%%%%%%%%%%%
\subsubsection{Compressible MHD shearing waves}
%
%
%
%%%%%%%%%%%%%%%%%%%%%%%%%%%%%%%%%%%%%%%%%%%%%%%%%%%%%%%%

\begin{figure}\centering
\includegraphics[width=0.5\textwidth]{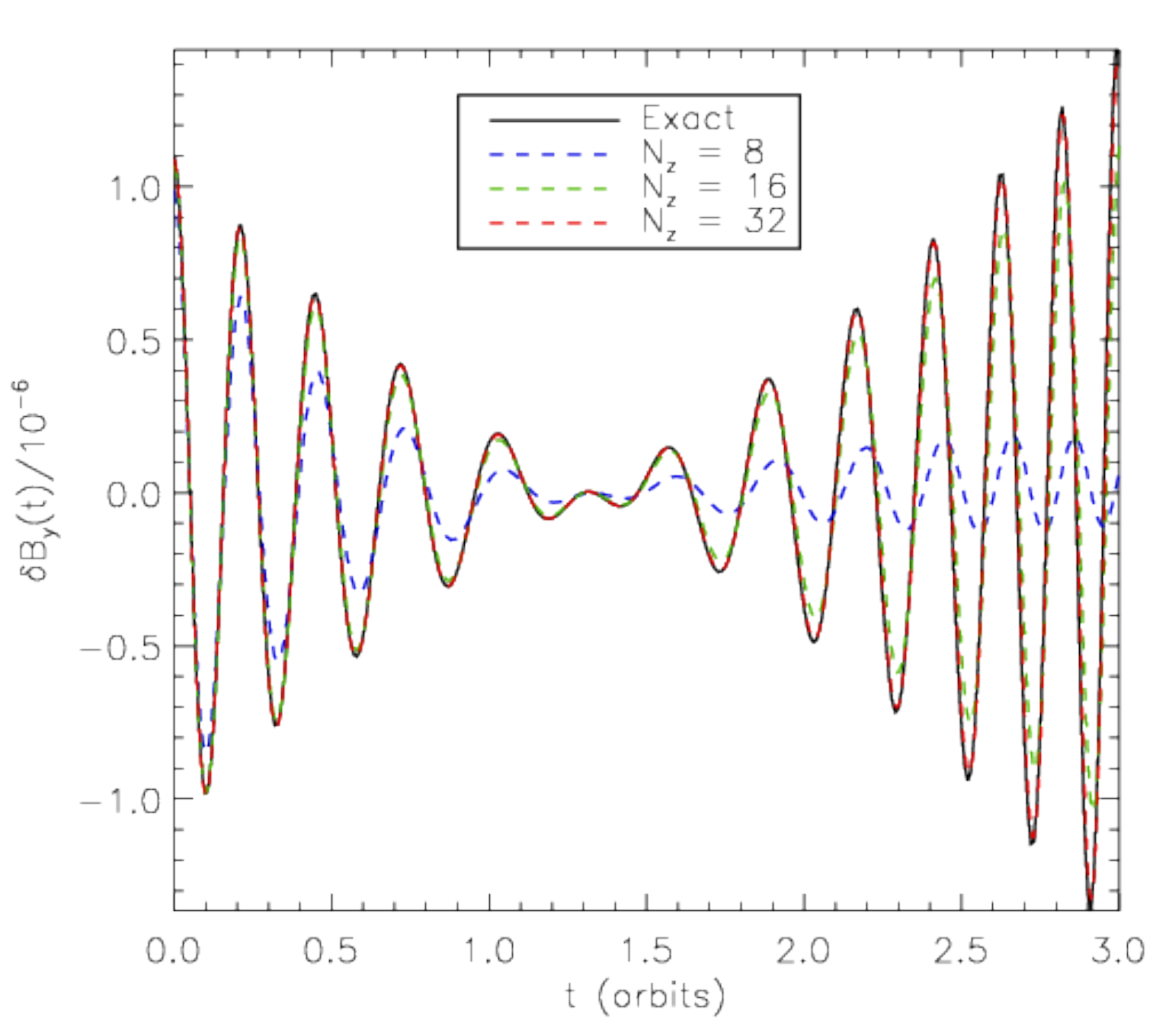}
\caption{\footnotesize Azimuthal magnetic-field perturbation as a function of time for the compressible MHD shearing-wave problem. The exact solution is plotted as a solid black line and the dashed lines correspond to computations obtained at the resolutions of $N_z=8$ (blue), $16$ (green), and $32$ (red) zones in the vertical direction.}
\label{fig:SHW}
\end{figure}
As a first benchmark, we consider the evolution of compressible, magnetohydrodynamical shearing-waves in an isothermal medium using the shearing-box model.
These waves can be regarded as the extension of fast and slow modes to a differentially rotating medium and provide the natural basis for wave decomposition in the linear theory of rotating MHD shear flows \citep{Johnson.2007}.

We follow the same configuration used by \citet{JohGuaGam.2008} and \citet{ StoGar.2010}  and consider a Cartesian domain of size $[-1/4,1/4]^3$ with  $N\times N/2\times N/2$ grid zones.
The initial conditions, the details of which may be found more precisely in \cite{JohGuaGam.2008-addendum}, consists of a background equilibrium state about which a plane wave perturbation is imposed
\begin{equation}\left\{\begin{array}{lcl}
 \rho &=& 1 + \delta\rho\cos\left(\vec{k}\cdot\vec{x}\right)
\\ \noalign{\medskip}
 \vec{v} &=& \DS \vec{w} + \frac{c_s}{2}\sqrt{\frac{7}{10}} \frac{H\vec{k}}{4\pi}\delta\rho
            \cos\left(\vec{k}\cdot\vec{x}\right)
\\ \noalign{\medskip}
  \vec{A} &=&\DS \left(0,0,\frac{y}{10}-\frac{x}{5}\right) 
             + \delta A \left(2,-1,5\right)\sin\left(\vec{k}\cdot\vec{x}\right)\,,
\end{array}\right.
\end{equation}
where $\vec{w}=-q\Omega_0x\hvec{y}$ is the background shear flow, $H=c_s/\Omega_0$ is the scale height, and $\vec{k}=4\pi/H(-2,1,1)$ is the initial wavenumber, whereas
\begin{equation}
 \delta\rho = \epsilon \left(8\pi\sqrt{\frac{10}{7}}\right)^{\HALF}
\,,\quad
 \delta A = \epsilon c_s \frac{H}{60}
            \left(\frac{1}{\pi}\sqrt{\frac{5}{14}}\right)^{\HALF}
\end{equation}
are the perturbation amplitudes of density and vector potential ($\epsilon=10^{-6}$). 
Magnetic field is initialized from $\vec{B}=\nabla\times\vec{A}$.
We set the isothermal sound speed to $c_s = 1$, while the orbital frequency is $\Omega_0 = 1$.

In Fig \ref{fig:SHW}, we compare, at different resolutions $N=8,16,32$, the temporal evolution of the azimuthal field perturbation
\begin{equation}
\delta B_y(t) = 2\int \left[B_y(t) - \hat{B}_y(t)\right]
                \cos\left(\vec{k}(t)\cdot\vec{x}\right)\, dV\,,
\end{equation}
where $\vec{k}(t) = \vec{k} + q\Omega k_y t\hvec{x}$ is the time-dependent wavenumber and $\hat{B}_y(t) = c_s(1/5 - q\Omega_0 t/10)$ is the analytical expectation, kindly provided to us by G. Mamatsashvili.
Our results are in excellent agreement with those of \cite{JohGuaGam.2008} and \cite{StoGar.2010} showing that at the highest resolution of $32$ zones per wavelength, the solution has converged to the semi-analytical prediction.
%

%%%%%%%%%%%%%%%%%%%%%%%%%%%%%%%%%%%%%%%%%%%%%%%
\subsubsection{Nonlinear MRI}
%
%
%%%%%%%%%%%%%%%%%%%%%%%%%%%%%%%%%%%%%%%%%%%%%%%

\begin{figure}[!ht]\centering
\includegraphics*[width=0.5\textwidth]{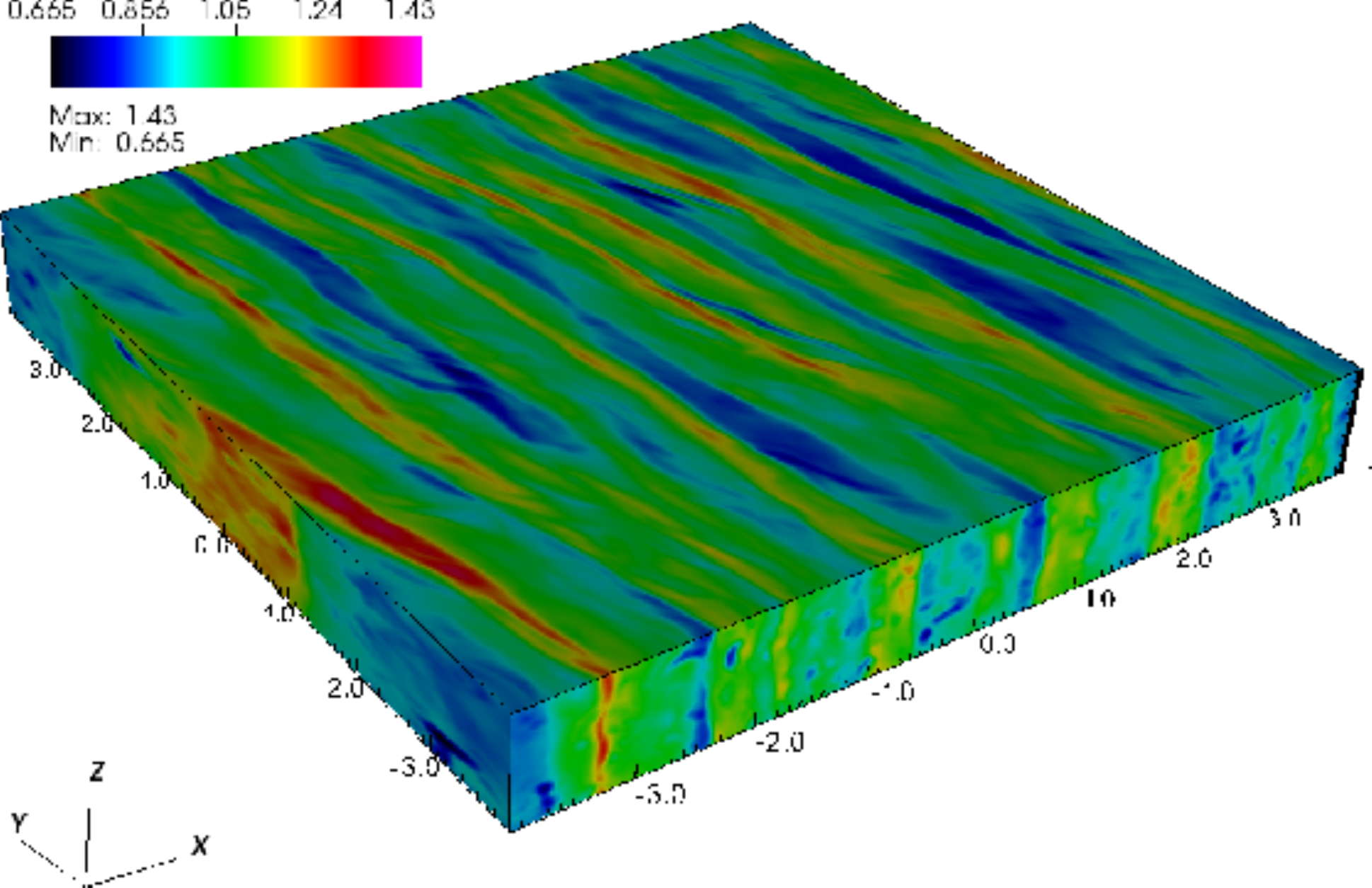}
\includegraphics*[width=0.5\textwidth]{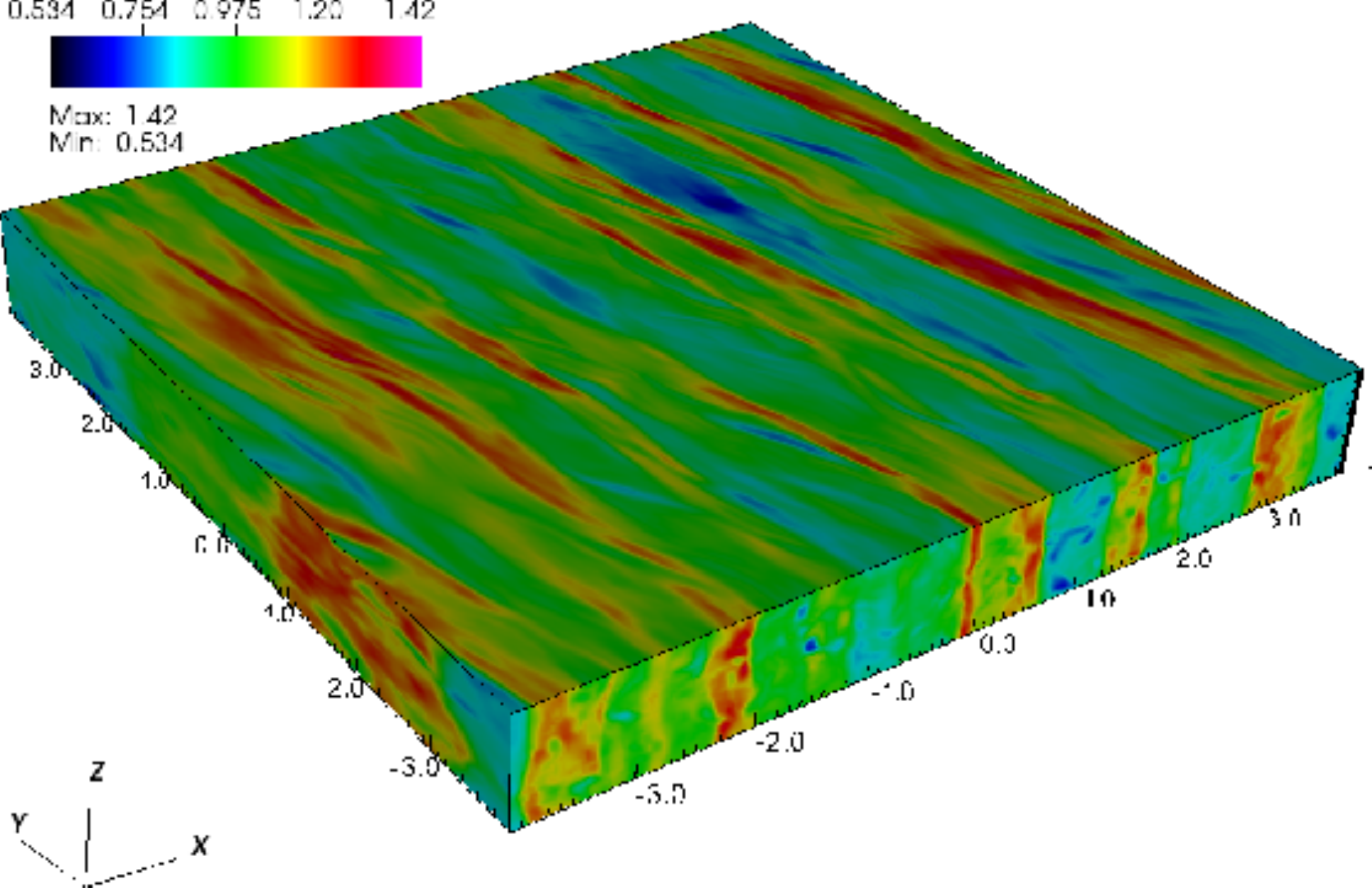}
\caption{\footnotesize Density distribution in the shearing-box computation
         after $40$ rotation periods. Top and bottom panel refers 
         to the model with and without FARGO-MHD, respectively.}
\label{fig:MRI-rho}
\end{figure}
\begin{figure}[!ht]\centering
\includegraphics[width=0.5\textwidth]{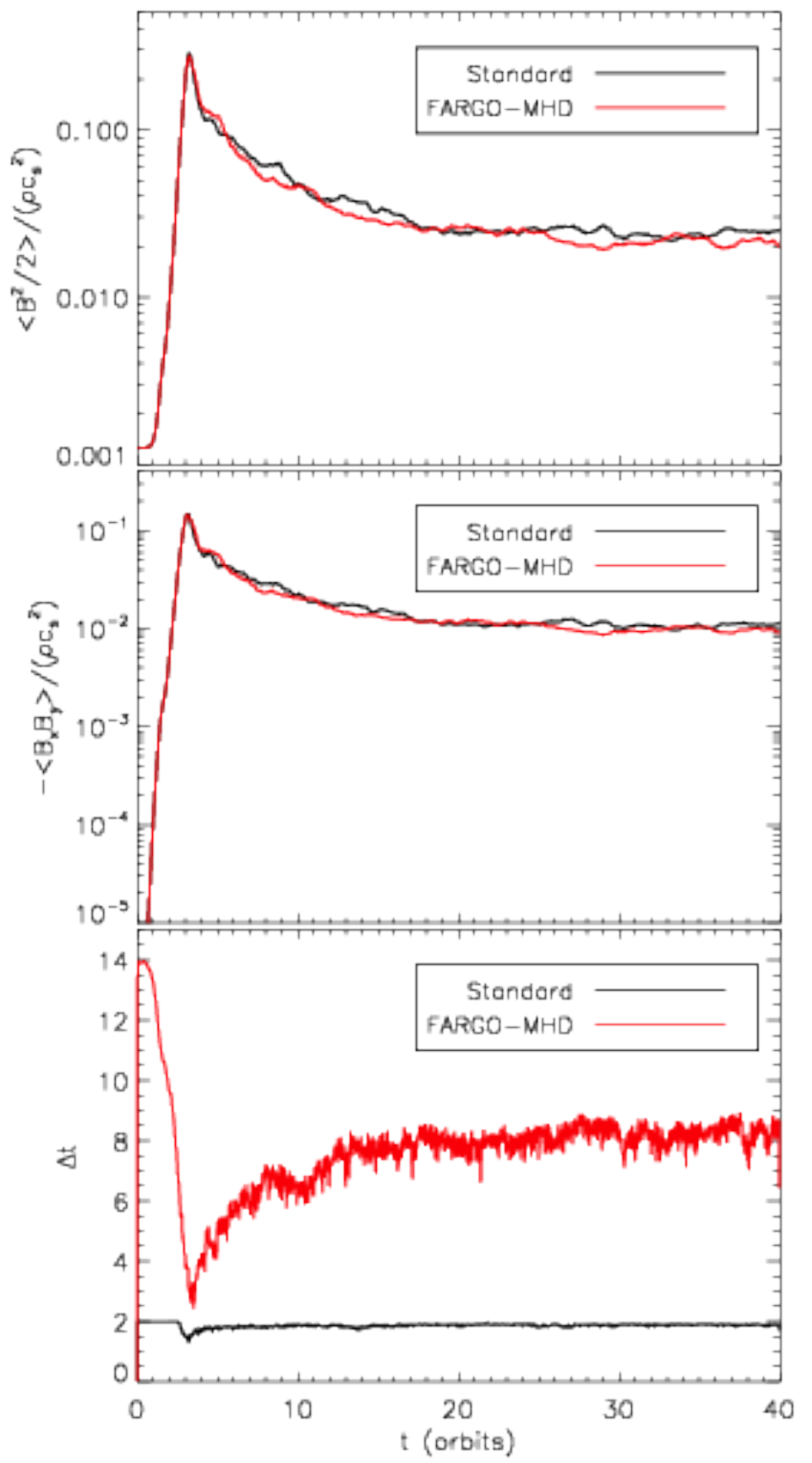}
\caption{\footnotesize Evolution of the MRI in the shearing-box model
         during the first $40$ orbits. 
         From top to bottom, the three panels show, respectively, the 
         volume-integrated magnetic energy-density, the Maxwell stresses,
         and the time increment used in the computations without
         (solid line) and with FARGO-MHD (dashed line).}
\label{fig:MRI-alpha}
\end{figure}

In the next example, we investigate the nonlinear evolution of the magneto-rotational instability (MRI) in the shearing-box model.
The initial condition consists of a uniform orbital motion in the $y$ direction $\vec{w}=-q\Omega_0 x\hvec{y}$ of constant density $\rho=1$ and zero net-flux magnetic field in the vertical direction
\begin{equation}
  \vec{B} = \sqrt{\frac{2\rho c_s^2}{\beta}}\sin\left(2\pi\frac{x}{8H}\right)
            \hvec{k}\,,
\end{equation}
where $\beta = 400$ and $c_s$ is the isothermal sound speed.
Following \citet{GreZie.2007}, \citet{JohGuaGam.2008}, and \citet{StoGar.2010}, we set $c_s=\Omega_0 = 10^{-3}$ so that the disk scale-height is $H=\Omega_0/c_s = 1$ and one orbital period corresponds to $2\pi/\Omega_0$ in code units.
The size of the computational domain, in units of $H$ is $[-4,4]\times[-4,4]\times[-\HALF,\HALF]$ and $32$ zones per scale-height are employed.
We assume periodic boundary conditions in the azimuthal ($y$) and vertical ($z$) directions, while shifted periodicity is imposed at the radial boundary.
We carry out two sets of computations, with and without orbital advection, at the resolution of $256\times 256\times 32$ zones.

The evolution is characterized by an initial transient phase where all the relevant physical quantities grow rapidly within a few orbits. 
Subsequently, the flow dynamics becomes nonlinear and the system settles down into a saturated regime accompanied by the formation of typical trailing spiral density waves as shown in Fig. \ref{fig:MRI-rho} at $t=80\pi/\Omega_0$ (i.e. after $40$ revolutions).
We note that, owing to the turbulent chaotic behavior of the MRI, a direct comparison between these structures is not really helpful but that one should instead inspect spatially averaged physical quantities.
Fig. \ref{fig:MRI-alpha} shows the temporal evolution of the volume-integrated magnetic energy density (top panel) and Maxwell stresses (middle) obtained with and without FARGO-MHD for the first $40$ orbits. 
The two computations are in excellent agreement with the absolute value of the (normalized) Maxwell stresses reaching, in both cases, the approximate constant value $\approx 0.01$.
The time step is plotted in the bottom panel of Fig \ref{fig:MRI-alpha}, where, after the initial transient phase, it is shown that the employment of orbital advection results in a value $\approx 4.3$ larger than the standard calculation, in accordance with the results of \cite{StoGar.2010}

%%%%%%%%%%%%%%%%%%%%%%%%%%%%%%%%%%%%%%%%%%%%%%%%%%%%%%%%%%%%%%%%%
\subsection{Disk-planet interaction}
%
%
%
%%%%%%%%%%%%%%%%%%%%%%%%%%%%%%%%%%%%%%%%%%%%%%%%%%%%%%%%%%%%%%%%%

\begin{figure}
\centering
\includegraphics*[width=0.45\textwidth]{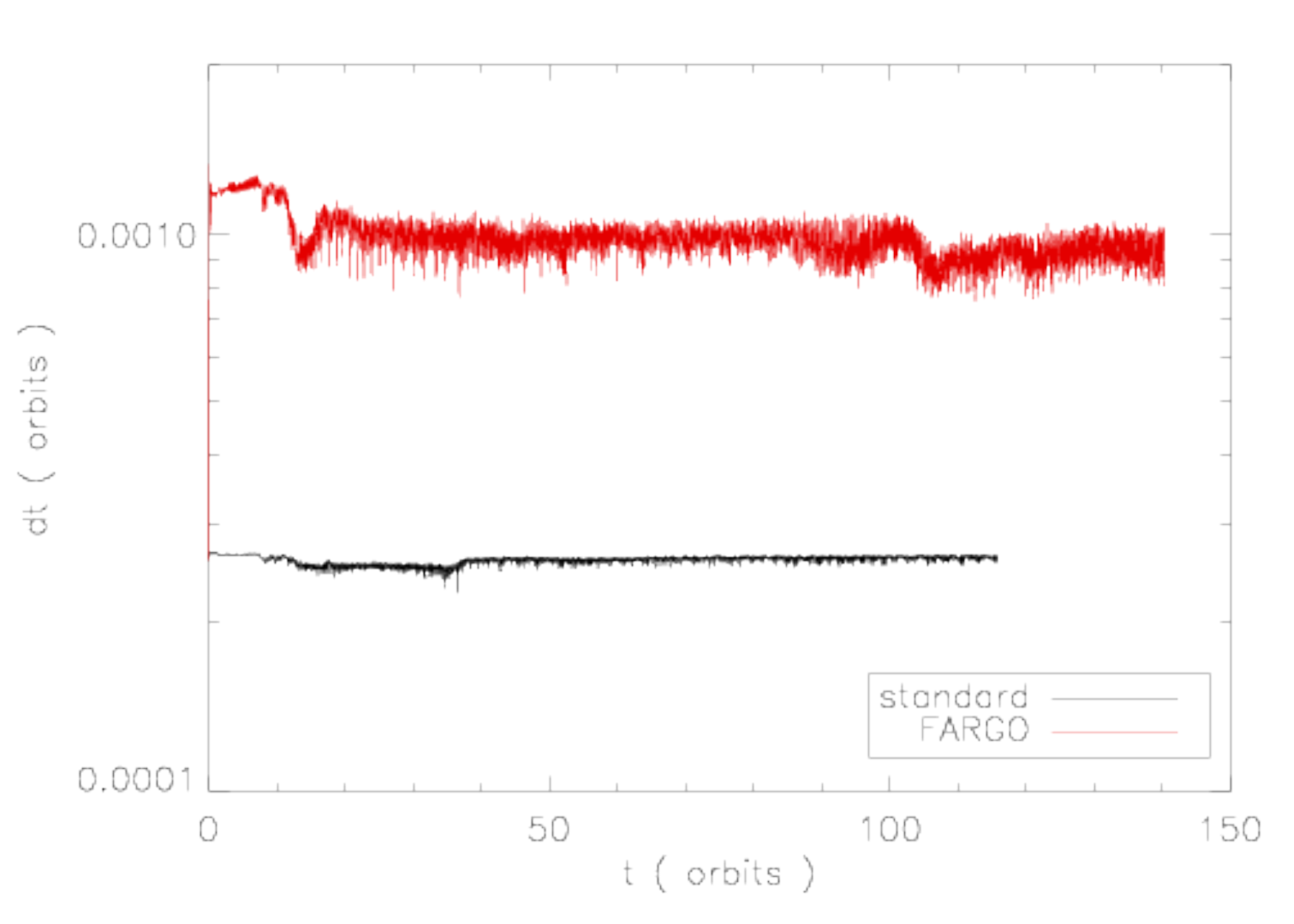}
\includegraphics*[width=0.45\textwidth]{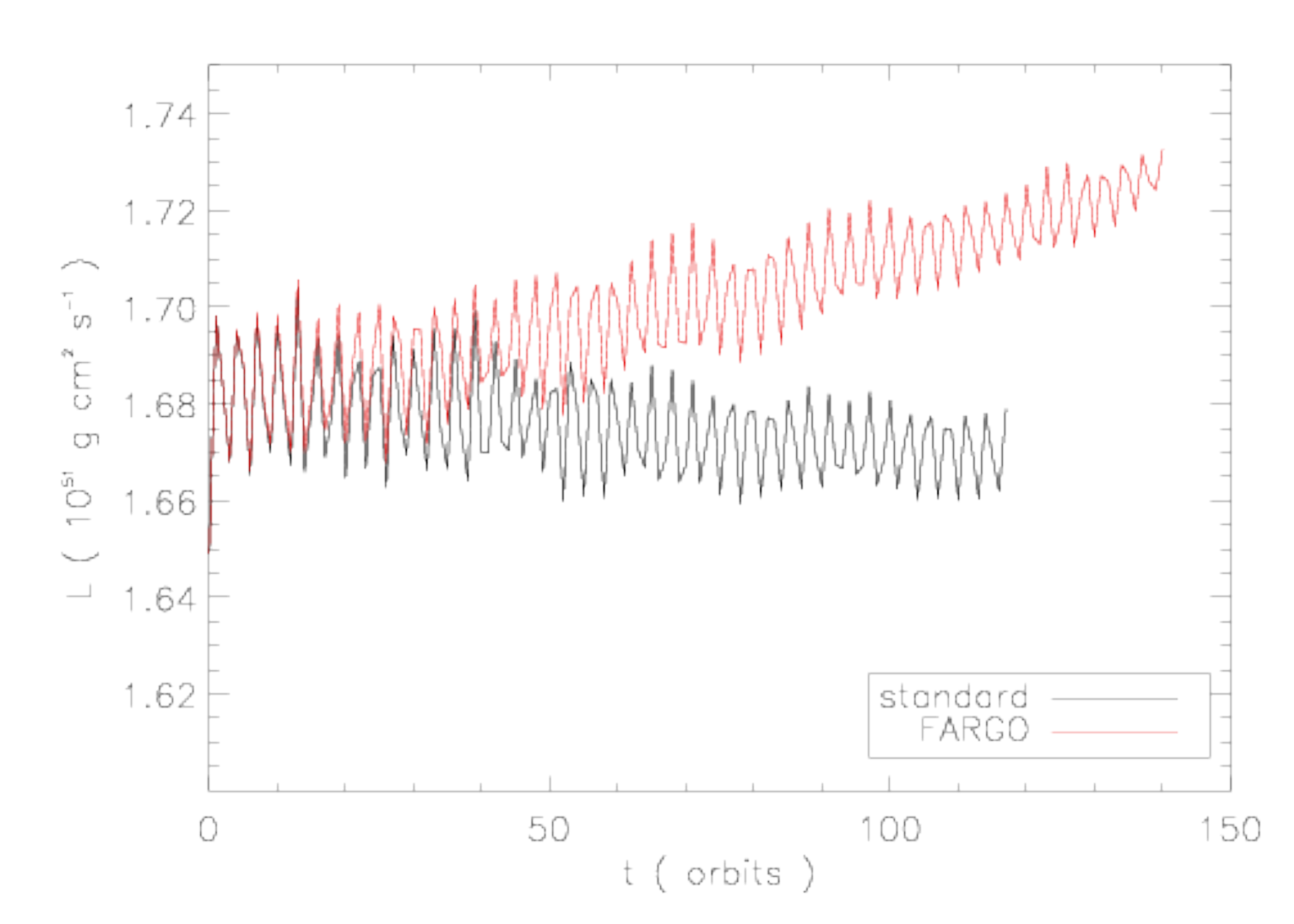}
\caption{Top: time step as a function of orbits for the disk-planet interaction problem using the standard scheme (black) and FARGO (red).
Bottom: change in angular momentum as a function of time.}
\label{fig:DiskPlanet_dt+L}  
\end{figure}
\begin{figure}[!ht]
\centering
\includegraphics[width=0.45\textwidth]{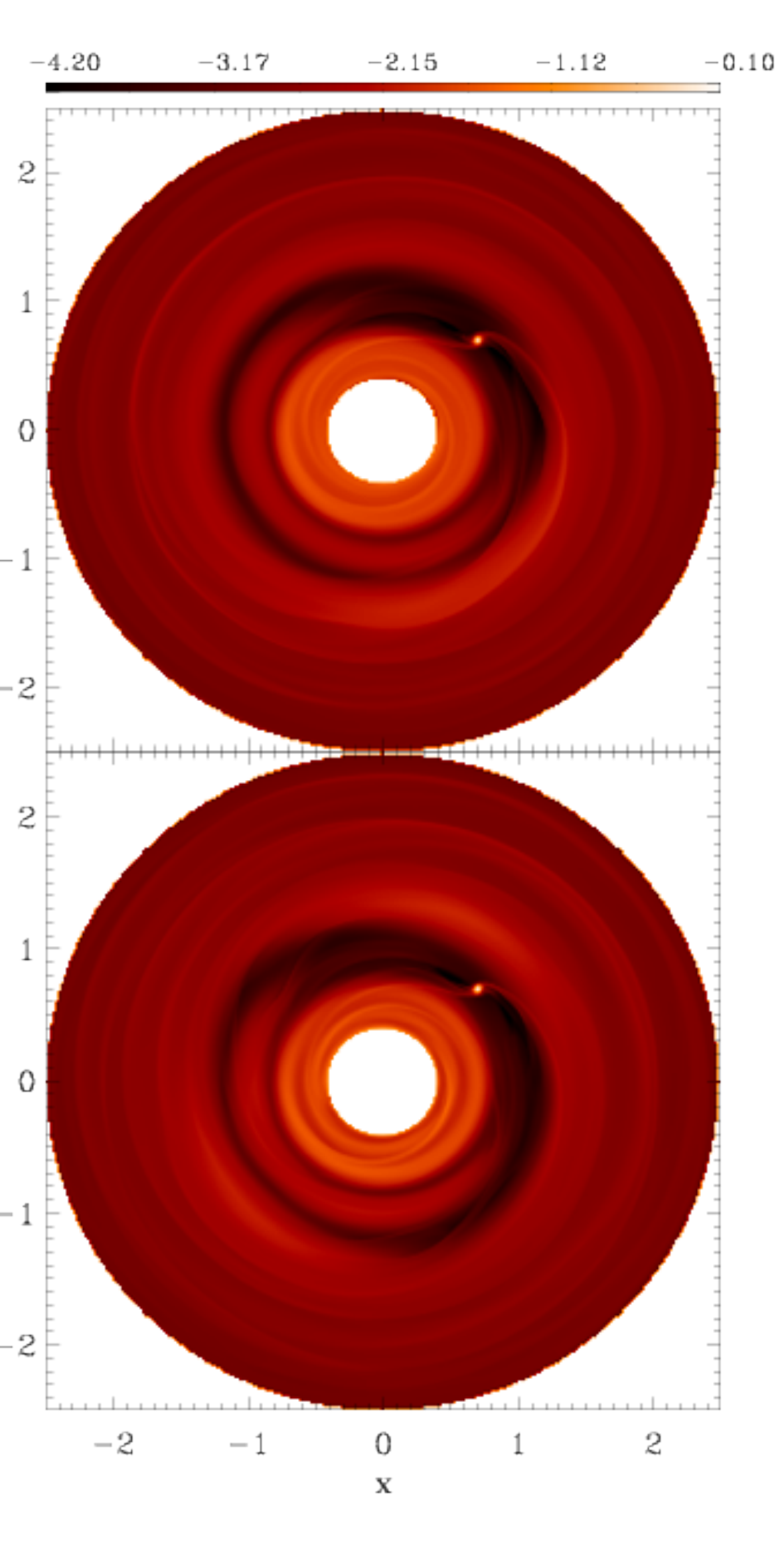}
\caption{Logarithmic density map for the disk-planet interaction after 100 orbits in the equatorial plane using FARGO (top) and the standard integration scheme (bottom).}
\label{fig:DiskPlanet_rho}  
\end{figure}

We simulate the interaction of a planet embedded in a viscous global disk as in \citet{KBK09}. 
The 3D ($r, \theta, \varphi$) computational domain consists of a complete annulus of the protoplanetary disk centered on the star, extending from $r_{\rm min} = 0.4$ to $r_{\rm max} = 2.5$ in units of $r_0 = a_{\rm Jup} = 5.2$ AU. 
In the vertical direction, the annulus extends from the disk's midplane (at $\theta = 90^\circ$) to about $7^\circ$ (or $\theta = 83^\circ$) above the midplane. 
The mass of the central star is one solar mass $M_* = M_\odot$ and the total disk mass inside $[r_{\rm min},r_{\rm max}]$ is $M_{\rm disk} = 0.01 M_\odot$. 
For the present study, we use a constant kinematic viscosity coefficient with a value of $\nu = 10^{15}$\,cm$^2$/s, which corresponds to an $\alpha$-value of $\alpha = 0.004$ at $r_0$ for a disk aspect ratio of $H = 0.05\,r\sin\theta$.
The resolution of our simulations is $(N_r , N_\theta , N_\varphi ) = (256, 32, 768)$. 
At the radial boundaries, the angular velocity is set to the Keplerian values, while we apply reflective, radial boundary-conditions for the remaining variables.
In the azimuthal direction we use periodic boundary conditions hold while zero-gradient is imposed at the vertical boundaries.

The models are calculated with a locally isothermal configuration where the temperature is constant on cylinders and has the profile $T(R) \propto R^{-1}$ with the cylindrical radius $R = r \sin \theta$. 
This yields a constant ratio of the disk's vertical height $H$ to the radius $R$, which was set to 0.05.  
The initial vertical density stratification is approximately given by a Gaussian
\begin{equation}
\rho(r,\theta)= \rho_0 (r) \, \exp \left[ - \frac{(\pi/2 - \theta)^2 \, 
r^2}{2\,H^2} \right].
\end{equation} 
Here, the density in the midplane is $\rho_0 (r) \propto r^{-1.5}$, which leads to a $\Sigma(r) \propto \, r^{-1/2}$ profile of the vertically integrated surface density. 
The vertical and radial velocities $v_\theta$ and $v_r$ are initialized to zero, while the initial azimuthal velocity $v_\varphi$ is given by the equilibrium between gravity, centrifugal acceleration, and the radial pressure gradient. 
The simulations are performed in a frame of reference co-rotating with the planet using a second-order Runge Kutta scheme with linear reconstruction and Courant number $C_a=0.25$.
Following \cite{Kley.1998}, we adopt a conservative treatment of the Coriolis force in the angular momentum equation.

The planet is described by adding a term to the total gravitational potential acting on the disk as 
\begin{equation}
\Phi  = \, \Phi_* + \Phi_{p} \,  = - \, \frac{G M_*}{r} - 
\frac{G m_{\rm p}}{\sqrt{({\bf r}-{\bf r}_{\rm p})^2}},
\end{equation}
where ${\bf r}_{\rm p}$ denotes the radius vector of the planet location. 
We follow \citet{KlK06} in using a cubic planetary potential 
\begin{equation}
\Phi_p =  \left\{
\begin{array}{cc} 
\DS - \frac{m_{\rm p}\, G}{d} \,  \left[ \left(\frac{d}{r_{\rm sm}}\right)^4
- 2 \left(\frac{d}{r_{\rm sm}}\right)^3 + 2 \frac{d}{r_{\rm sm}}  \right]
\quad &  \mbox{for} \quad  d \leq r_{\rm sm}  \\ \noalign{\medskip}
\DS -  \frac{m_p G}{d}  \quad & \mbox{for} \quad  d > r_{\rm sm} \quad,
\end{array}
\right.
\end{equation}
where $r_{\rm sm}= 0.6H$.
We compute our model using both the standard integration method and the proposed FARGO-MHD scheme (without magnetic field).
The employment of the FARGO scheme increased the time-step by a factor of about 3.8 (top panel in Fig. \ref{fig:DiskPlanet_dt+L}). 

%Owing to our choice of boundary conditions and the presence of the planet, we do not fully conserve the total angular momentum, both dominated by the azimuthal velocity component, but the change of both quantities with time is in good agreement in both runs (bottom panel in Fig. \ref{fig:DiskPlanet_dt+L}).

Owing to our choice of boundary conditions and the presence of the planet, we do not fully conserve the total angular momentum although the angular momentum fluctuations in time, computed with or without orbital advection, are in good agreement (bottom panel in Fig. \ref{fig:DiskPlanet_dt+L}).

\begin{figure}
\centering
\includegraphics[width=0.45\textwidth]{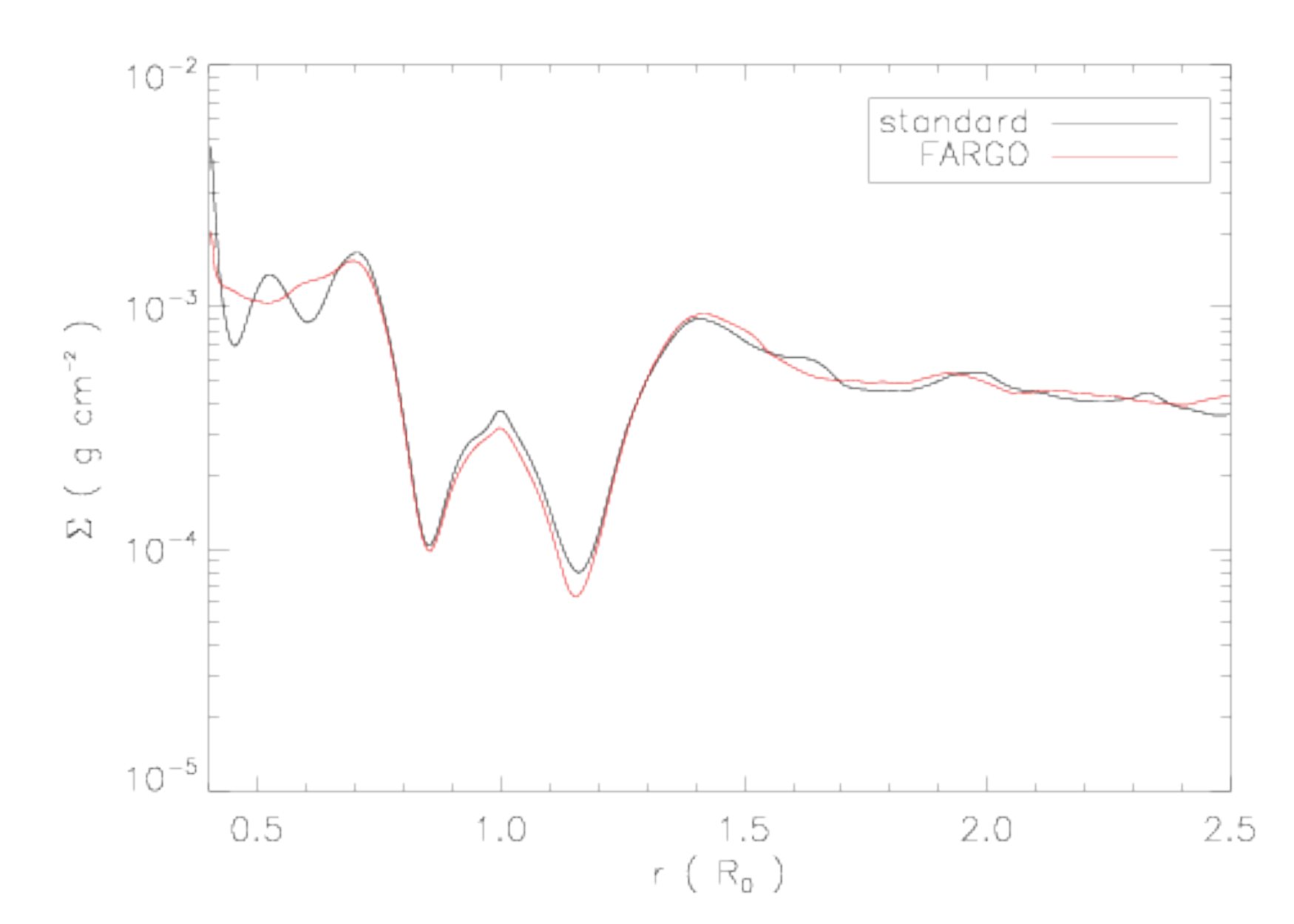}
\includegraphics[width=0.45\textwidth]{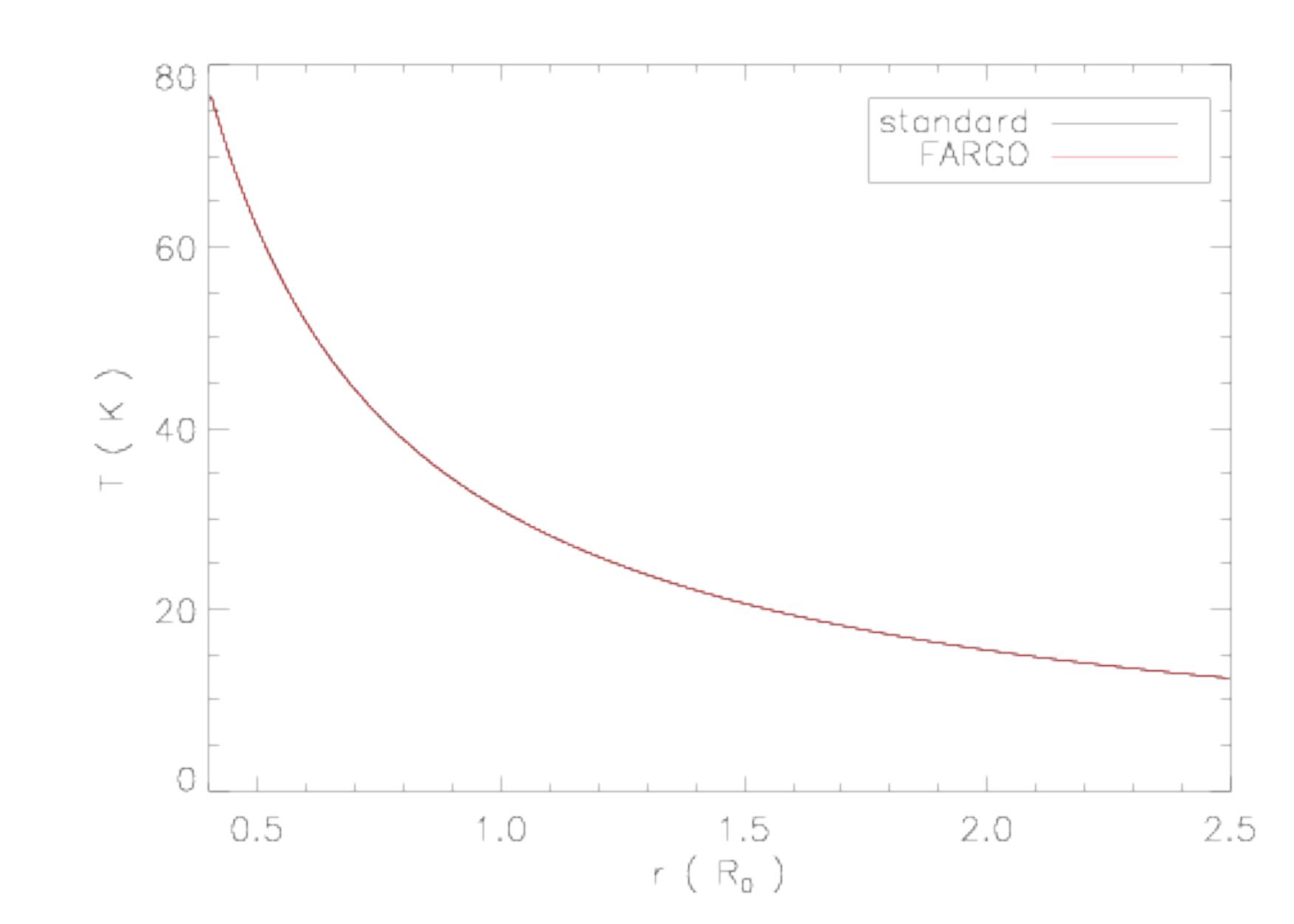}
\caption{Profiles of surface density (top) and temperature (bottom) as a 
function of radius for the disk-planet interaction problem. Solid red and black lines refer to computations obtained using FARGO and the standard integration method, respectively.} 
\label{fig:DiskPlanet_profiles}  
\end{figure}

In Fig. \ref{fig:DiskPlanet_rho}, we show logarithmic maps of density in the equatorial plane after 100 orbits comparing the computations obtained with and without orbital advection. The differences are only marginal. 
As a more precise test, we also plot the profiles of surface density and temperature as a function of radius, averaged over the azimuthal direction (Fig. \ref{fig:DiskPlanet_profiles}). 
The global slope of the profiles as well as the position and width of the gap created by the planet are in excellent agreement between both runs. 

%%%%%%%%%%%%%%%%%%%%%%%%%%%%%%%%%%%%%%%%%%%%%%%
\subsection{Linear MRI in global disk}
%
%
%%%%%%%%%%%%%%%%%%%%%%%%%%%%%%%%%%%%%%%%%%%%%%%

We test the FARGO-MHD scheme on the linear MRI in global disks, following the setup presented by \cite{Flock_etal.2010}.
This test is very sensitive to the numerical consistency and, at the same time, the MRI growth rates are affected by the numerical dissipation of the scheme. 

An analytical description of the linear stage of MRI was given by \cite{BalHaw.1991}, while global simulations as well as the nonlinear evolution of the MRI was presented in \cite{HawBal.1991}.
The absolute limit of the growth rate for ideal MHD is given for the zero radial wave-vectors $q_r = 0$ with the normalized wave vector $q_z$
\begin{equation}
q_z=k_z\sqrt{16/15}V_A/\Omega.
\end{equation} 
with the Alfv\'en velocity $V_A=B_z/\sqrt{4\pi\rho}$.
The critical mode $q_z=0.97$ grows exponentially ($\Psi = \Psi_0 e^{\gamma t}$) with growth rate $\gamma = 0.75 \Omega$.
For global disk models with disk thickness $H$, the critical wavelength can be rewritten to
\begin{equation}
\frac{\lambda_{\rm crit}}{2H} 
 = \sqrt{\frac{16}{15} } \frac{\pi V_{\rm A}}{\Omega H}
\end{equation}
with the angular frequency $\Omega = R^{-1.5} $ and the Alfv\'en velocity $V_{\rm A}$ \citep[see also Eq. 2.3 in][]{HawBal.1991}.

We use polar cylindrical coordinates $(R,\phi, Z)$ with uniform resolution in the region $0\le\phi\le\pi/3$, $-R_0/2\le z \le R_0/2$, and $R_0\le R\le4R_0$, where $R_0$ is the unit length.   
The initial density $\rho$ and pressure $p$ are constant across the entire disk patch with $\rho=1.0$, $P=c_s^{2}\rho/\Gamma$, $c_s=0.1V_{\phi,0}$, and $\Gamma = 1.00001$. 
The gas is initially set up with a Keplerian speed $V_{\phi,0}^2=R_0/R$.
A uniform vertical magnetic-field is placed in the region $2R_0\le R\le 3R_0$.
We choose the strength of the vertical magnetic field to obtain the four fastest-growing modes, fitting in the domain at $R=2$, $B_{z}=B_0/n$ where $B_0=0.055$ and $\rm n=4$.
We use three different resolutions $[R,\phi,Z]=[224,168,64],[112,84,32],[66,42,16]$ with a logarithmic increasing grid to measure the capability to resolve the MRI wavelength with $16$, $8$, and $4$ grid cells per vertical scale-height, respectively.
We choose $V_{R}(z)=V_{0}\sin{4z/H}$  for initial radial velocity 
with the vertical size $H$ of the box.
Boundary conditions are periodic for all variables in the vertical and azimuthal directions, while zero gradient is imposed on all flow quantities at the radial boundaries.

We employ the HLLD Riemann solver \cite{MiyKus.2005}, piece-wise linear reconstruction and the second-order Runge Kutta time-stepping scheme with and without FARGO-MHD. 
The Courant number is set to $0.33$ and the spatial reconstruction of the electromotive force at the zone edges is carried out using the approach described in \cite{GarSto.2005}

\begin{table}
\caption{Growth rates of the linear MRI mode.}
\label{tab:linearMRI}
\centering
\begin{tabular}{ccccc}\hline
$N_z$/mode & $\gamma^{MRI}$ (std $2^{nd})$ & 
$\gamma^{MRI}$ (Fargo $2^{nd})$ &
$\gamma^{MRI}$ (Fargo $3^{rd})$  \\
\hline
\hline
16 & 0.70  & 0.71 & 0.74 \\
8 & 0.57 & 0.59 & 0.69  \\
4   & 0.00  & 0.14 & 0.14  \\
\hline
\end{tabular}
\tablefoot{\footnotesize The first column gives the resolution per wavelength, while columns 2 to 4 show, respectively, the MRI growth rates obtained using the standard scheme, FARGO-MHD with linear reconstruction (Eq. \ref{eq:LTS_MH_flux}) and FARGO-MHD with parabolic reconstruction (PPM, Eq. \ref{eq:LTS_PPM_flux}).}
\end{table}

In Table \ref{tab:linearMRI}, we present the growth of radial magnetic energy over radius during the linear MRI phase using the standard scheme, as well as FARGO-MHD, with piece-wise linear (Equation \ref{eq:LTS_MH_flux}) and piece-wise parabolic reconstruction (Equation \ref{eq:LTS_PPM_flux}). 
We determine the growth rate from the time derivative of the amplitude maxima for $B_R$ in Fourier space at each radius.
Computations obtained with the FARGO-MHD scheme provides, at the same resolution, a higher growth rate than the standard scheme.
This trend has to be attributed to the reduced numerical dissipation as confirmed by the progressive increase in the growth rate when moving from a second-order to a third-order interpolation scheme.
The observed speedup is around $4$.
We also show that we need at least 8 grid cells per wavelength to correctly resolve the growth rate. 
The third-order reconstruction method in combination with FARGO-MHD reaches with 16 grid cells per H, growth rates up to the analytical limit of 0.75. 

%%%%%%%%%%%%%%%%%%%%%%%%%%%%%%%%%%%%%%%%%%%%%%%
\subsection{Turbulent accretion disk}
%
%
%%%%%%%%%%%%%%%%%%%%%%%%%%%%%%%%%%%%%%%%%%%%%%%

\begin{figure*}[!ht]\centering
\includegraphics*[width=0.45\textwidth]{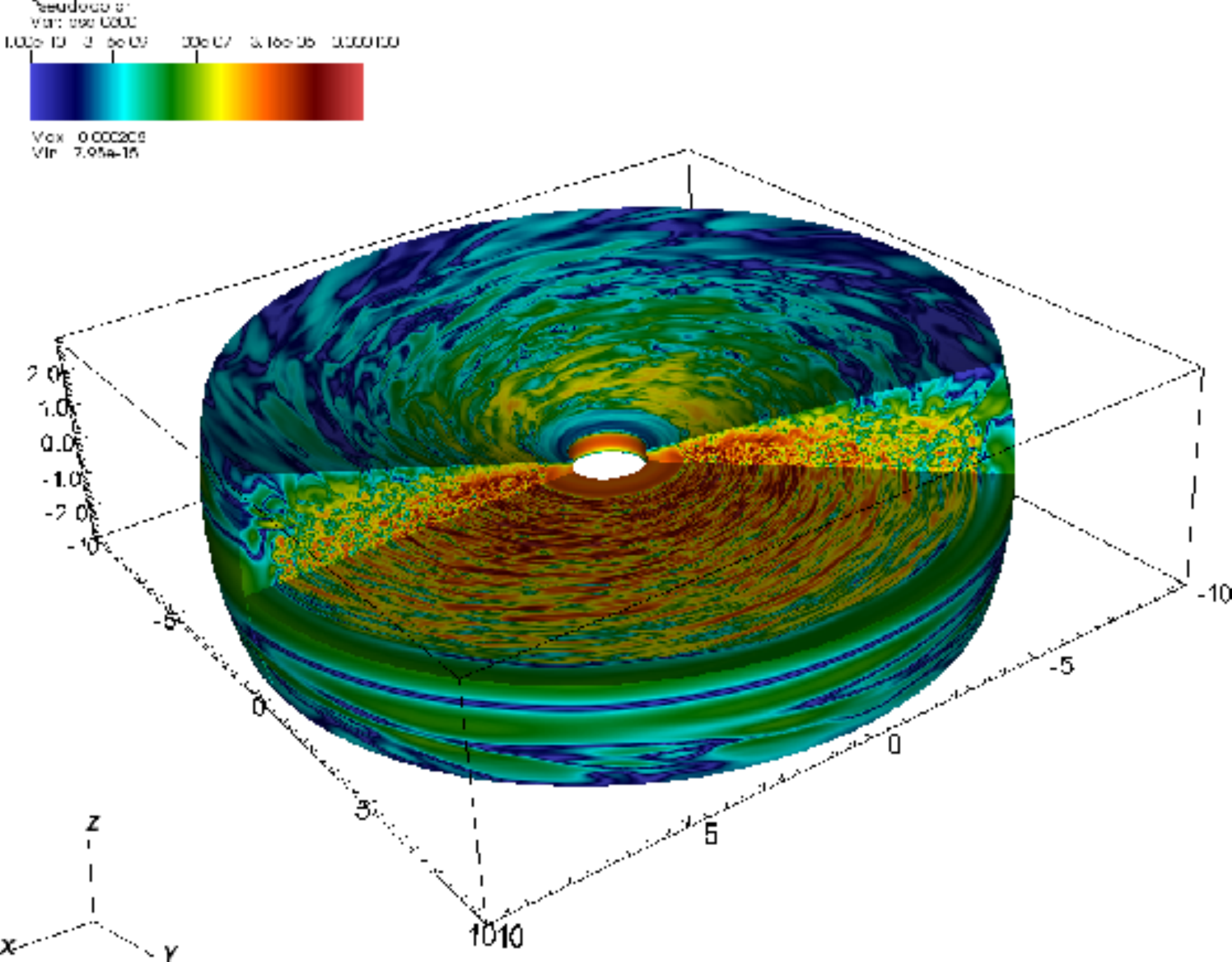}%
\includegraphics*[width=0.45\textwidth]{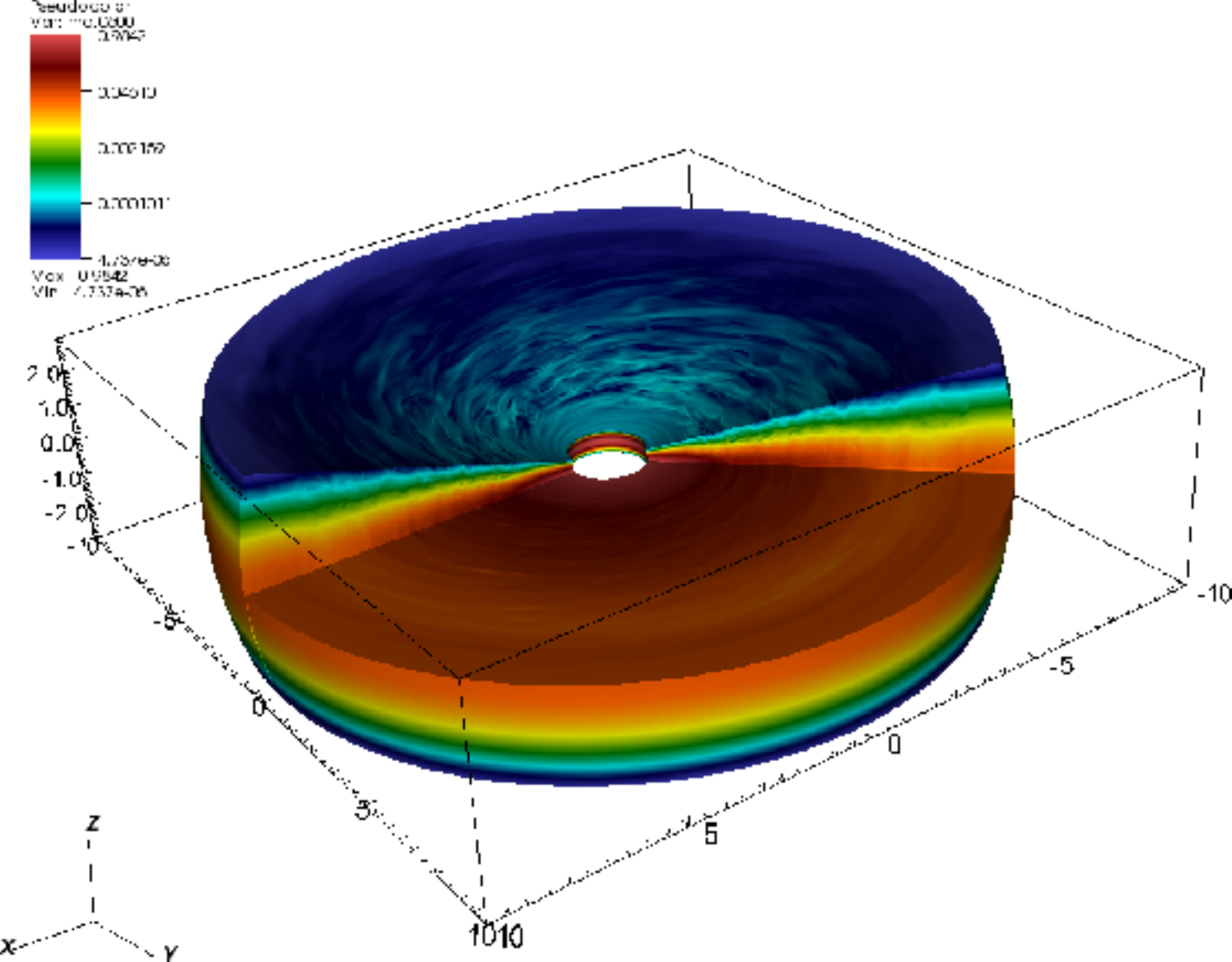}
\caption{\footnotesize Three-dimensional cross-sections of the vertically stratified magnetized accretion disk at $t=100$ for the high-resolution run showing the surface map of the magnetic pressure $\vec{B}^2$ (left) and density (right).}
\label{fig:Disk3D}
\end{figure*}

As a final application, we consider a turbulent accretion disk in 3D spherical coordinates $(r,\theta,\phi)$ using the setup BO described by \cite{Flock_etal.2011}.
The initial condition consists of a vertically stratified structure in a Newtonian central potential $\Phi = -1/r$ with density 
\begin{equation}
 \rho = \rho_{0} R^{-3/2}
        \exp\left(\frac{\sin{(\theta)}-1}{(H/R)^2}\right) \,,
\end{equation}
where $\rho_{0} = 1$, $R = r \sin{(\theta)}$ is the cylindrical radius, and $\rm H/R = c_0 = 0.07$ defines the ratio of scale-height to radius.
The pressure follows locally the isothermal equation of state $p = c_{s}^2\rho$ where the sound speed $\rm c_{s} = c_0/\sqrt{R}$.
The azimuthal velocity is set to 
\begin{equation}
 V_{\phi} = 
 \sqrt{\frac{1}{r}}\left(1- \frac{2.5}{\sin(\theta)}c^2_0 \right).
\end{equation}
The initial velocities $V_{r}$ and $V_{\theta}$ are set to a white noise perturbation with amplitude $10^{-4} c_{s}$.
The simulation uses a pure toroidal seed magnetic field with constant plasma $\beta = 2P / B^{2} = 25$.

For the sake of comparison, we employ the same computational domain, numerical resolution, and boundary conditions adopted by \cite{Flock_etal.2011}. 
We thus have $1\le r\le 10$ (in AU), $\pi/2 - 0.3\le\theta\le\pi/2 + 0.3$ (approximately $\pm 4.3$ disk scale-heights), and $0\le\phi\le 2\pi$.
%with an aspect ratio at 5 AU of $1:0.67:1.74$ $(\Delta r:
%r\Delta\theta:r\Delta\phi\sin{\theta})$. 
The grid resolution is uniform with $N_r= 384$, $N_\theta=192$, and $N_\phi=768$ zones in the three directions.
Buffer zones extend from 1 AU to 2 AU as well as from 9 AU to 10 AU.
In the buffer zones, we use a linearly increasing resistivity (up to $\eta = 10^{-3}$) reaching the boundary. 
This dampens magnetic field fluctuations and suppresses the interactions with the boundary. 
Our outflow boundary condition projects the radial gradients in density, pressure, and azimuthal velocity into the radial boundary, and the vertical gradients in density and pressure at the $\theta$ boundary. 
For the first run, we employ the HLLD Riemann solver, piece-wise linear reconstruction and $2^{nd}$ order Runge Kutta time integration. 
The second run is repeated with the FARGO-MHD scheme described in this paper. 

As the evolution proceeds, the initial configuration becomes vulnerable to the MRI instability, which quickly leads to a turbulent behavior.
Fig. \ref{fig:Disk3D} shows a slice-cut of the magnetic pressure and density after $600$ orbits.

\begin{figure}
\centering
\includegraphics[width=0.4\textwidth]{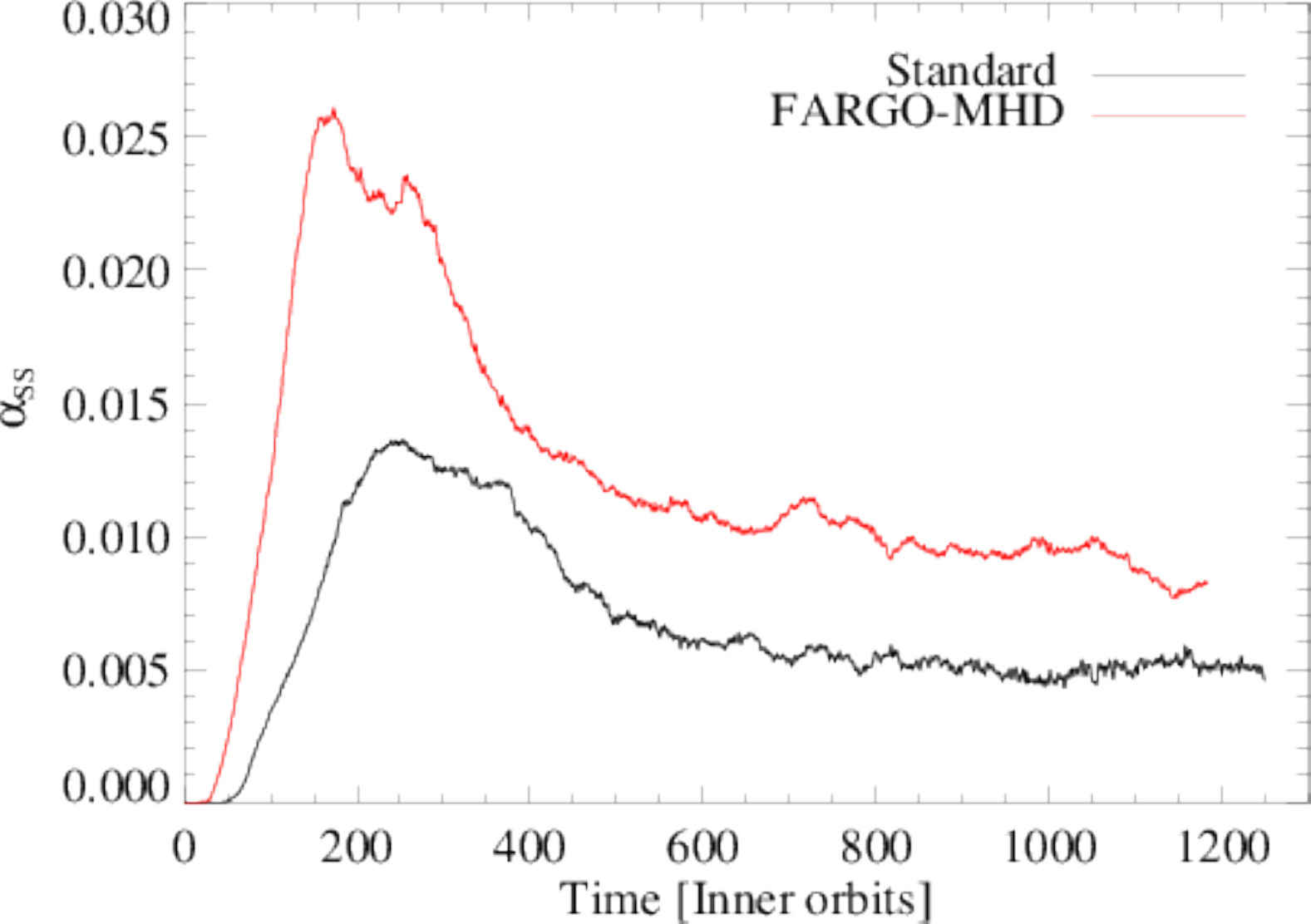}
\includegraphics[width=0.4\textwidth]{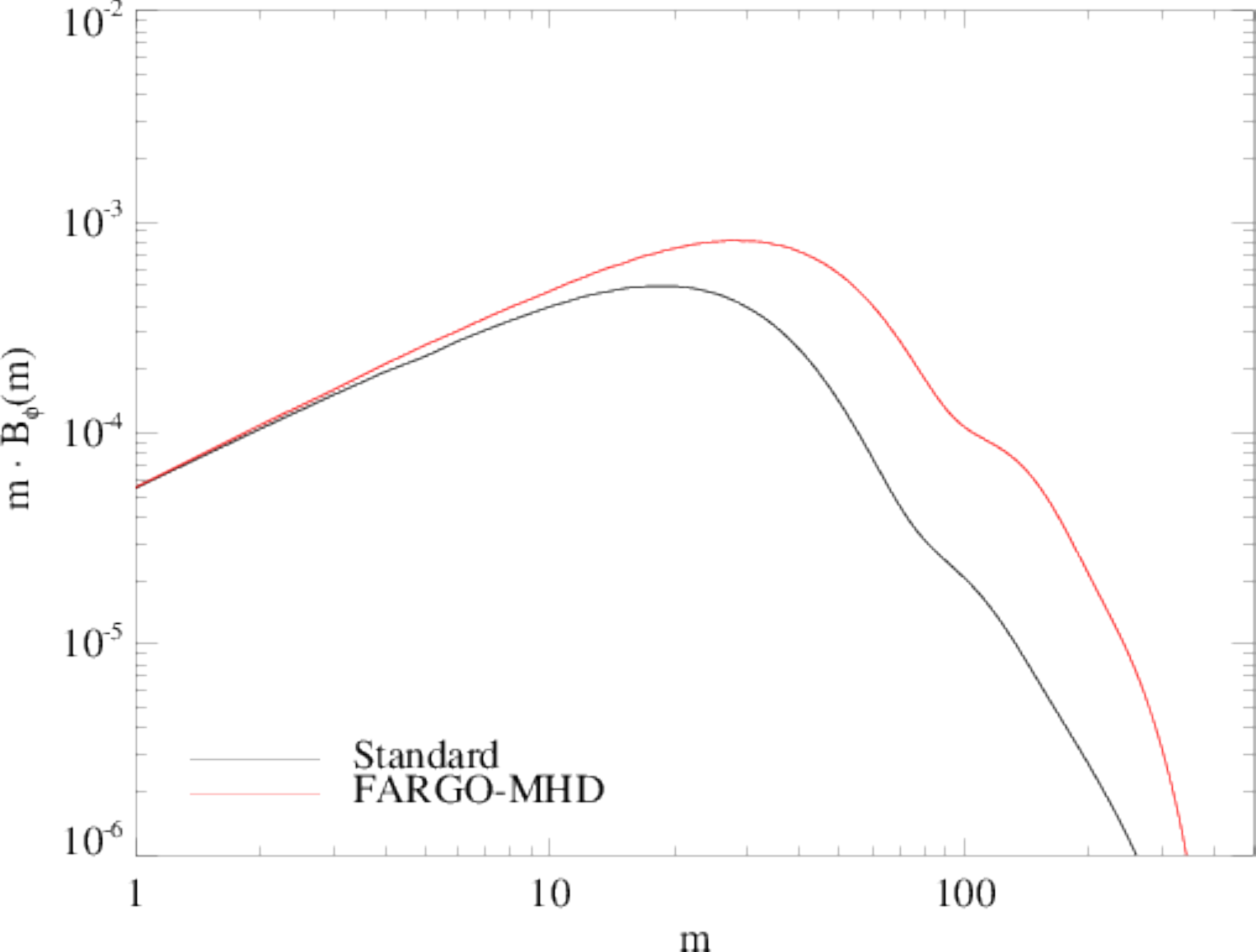}
\includegraphics[width=0.4\textwidth]{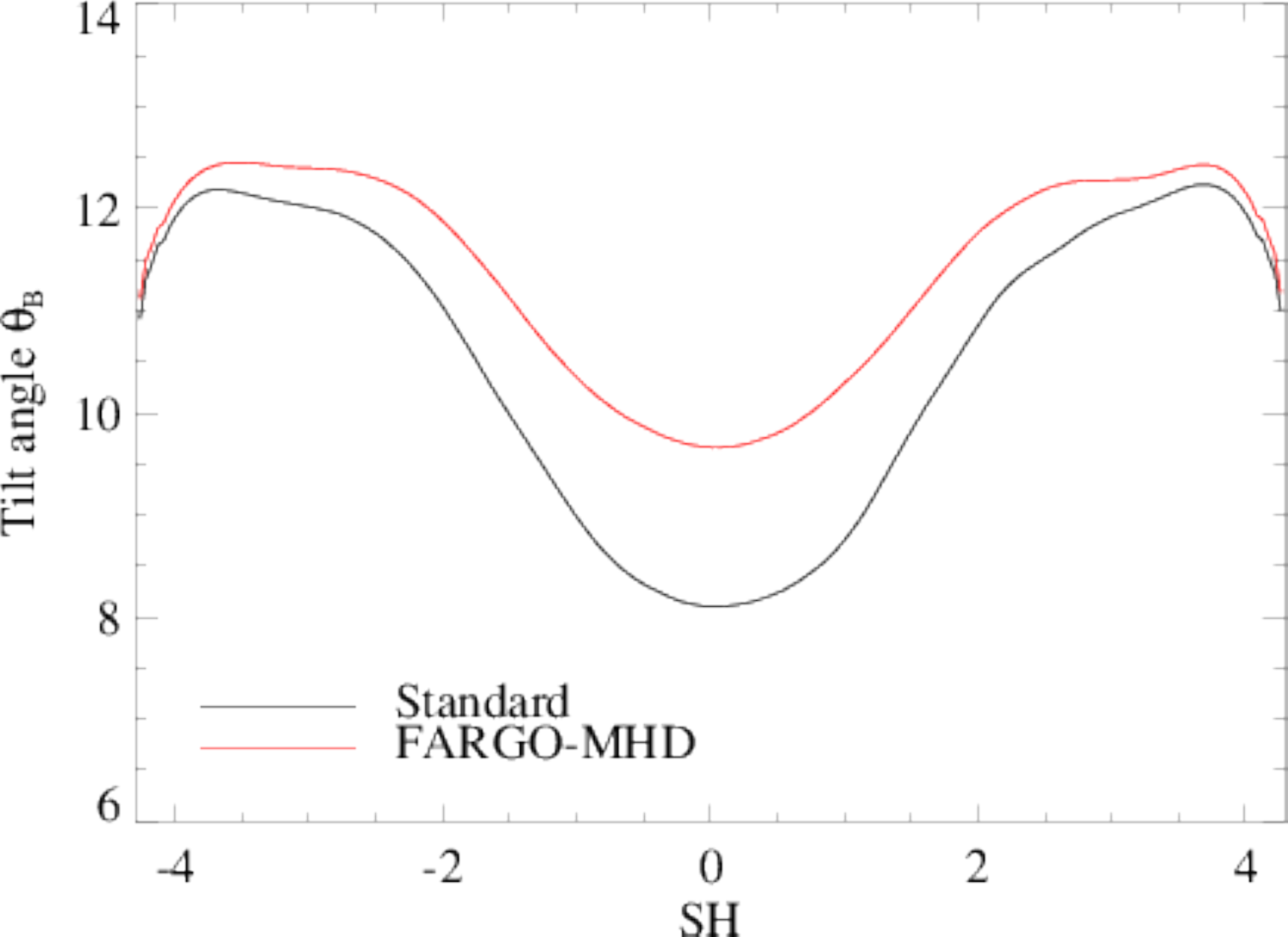}
\caption{Top: Time history of the accretion stress in the 
         turbulent accretion disk problem. Middle: Magnetic field spectra over azimuth. The FARGO method resolves much smaller turbulent scales using the same resolution. Bottom:Magnetic tilt angle over height.}
\label{fig:Disk3D-averages}
\end{figure}

A direct comparison between the results obtained with the FARGO-MHD scheme and those of \cite{Flock_etal.2011} is provided in the three panels of Fig \ref{fig:Disk3D-averages} by plotting relevant volume- and time-averaged quantities. 
For our analysis, we use the range between 3 AU and 8 AU (in $r$), which is unaffected by the buffer zones. Temporal averages are taken between 
$600$ and $1200$ inner orbits.
A detailed description of the measurement can be found in Section 2.1 in \citet{Flock_etal.2011}.

The volume-averaged values $\av{\alpha_{SS}}$ of the Shakura and Sunyaev parameter defined by
\begin{equation}
\rm \alpha_{SS} = \frac{ \int \rho \left(
\frac{v'_{\phi}v'_{R}}{c^2_s} -
\frac{B_{\phi}B_{R}}{4 \pi \rho c^2_s}\right)dV} {\int \rho dV}
\end{equation}
are plotted, as a function of time, in the top panel of Fig \ref{fig:Disk3D-averages}.
The FARGO-MHD run shows a much faster increase in the stresses even within the first 200 inner orbits, at the linear phase of MRI. 
While the standard computation shows a time-averaged $\alpha_{SS}$ value of $5 \cdot 10^{-3}$, the results obtained with FARGO-MHD reveal a factor of $\approx 2$ increase ($\alpha_{SS} = 9.6 \cdot 10^{-3}$).

The spectra of $B_{\phi}(m)$ over azimuthal wavenumber $m$, plotted in the middle panel of Fig. \ref{fig:Disk3D-averages}, reveals smaller resolved scales (within a factor of two) of the turbulent magnetic field. 
%There is roughly a factor of 2 smaller resolved scales.

The unstratified global simulations of \citet{Sorathia_etal.2011} suggest that the magnetic tilt angle could be a reliable indicator of numerically resolved MRI.
We measure the tilt angle for the magnetic field, which is defined by $\sin{2\theta_B} = |B_rB_\phi|/B^2$, at 4.5 AU and plot the time-averaged vertical profile of $\theta_B$ in the bottom panel of Fig. \ref{fig:Disk3D-averages}.
The tilt angle present the highest values in the coronal region. 
Employment of the proposed orbital-advection scheme results in a overall larger tilt angle, of $\approx 10^\circ$ in the midplane, compared to the standard computation, where $\approx 8^\circ$ in the midplane. 
These results suggest that, at the same resolution, the MRI is more accurately resolved when using orbital advection.

With the given grid resolution, chosen to match the results of \cite{Flock_etal.2011}, we observe a time-step speedup of $\sim 3.75$ when employing FARGO-MHD.
Nonetheless, we note that, according to the guidelines given in \S\ref{sec:FARGO}, this choice may not be an optimal one since the cell aspect ratio is $1:0.67:1.74$ ($\Delta r:r\Delta\theta:r\Delta\phi$), and larger gains may be obtained by suitably re-adjusting the number of points in order to have cells with aspect ratios closer to one.

%%%%%%%%%%%%%%%%%%%%%%%%%%%%%%%%%%%%%%%%%%%%%%%%%%%%%%%%%%%%%%%
\section{Summary}
\label{sec:summary}
%
%
%
%
%
%
%%%%%%%%%%%%%%%%%%%%%%%%%%%%%%%%%%%%%%%%%%%%%%%%%%%%%%%%%%%%%%%

We have presented an orbital advection scheme suitable for the numerical simulations of magnetized, differentially rotating flows in different systems of coordinates.
The algorithm has been implemented in the release 4.0 of the PLUTO code for astrophysical fluid dynamics \citep{PLUTO.2007,PLUTO.2011} and shares the same ideas as proposed by \cite{Masset.2000}, which consist of decomposing, at each time-step, the total velocity into an azimuthally averaged, mean contribution and a residual term. 
By taking advantage of operator splitting, the two contributions are carried out as a regular update of the standard MHD equations written in the residual velocity and a linear transport step corresponding to a non-integer shift of zones.
During the former, any dimensionally unsplit scheme may be employed provided that additional source terms are taken into account and correctly discretized to preserve conservation of both total angular momentum and total energy to machine precision.
In both steps, the magnetic field is evolved using the constrained transport formalism to maintain the divergence-free condition.

This approach yields substantially larger time steps whenever the orbital speed exceeds any other characteristic wave signal since the Courant condition depends on the residual velocity rather than the total velocity.
Numerical tests confirm that the proposed orbital advection scheme yields results that are equally accurate to and less dissipative than the standard numerical approach at a reduced numerical cost. 
The overall gain is problem-dependent and can be optimized by a suitable choice of grid resolution in the specified geometry: for the selected problems, we observed speedup factors between $4$ and $12$.

Parallel-domain decomposition can be performed in all three coordinate directions thus allowing the algorithm to be efficiently employed on large numbers of processors on modern parallel computers.
The proposed FARGO-MHD scheme is particularly suited to large-scale global disk simulations that have only recently become amenable to numerical computations on petascale systems.

\begin{acknowledgements}
This work has been supported by the PRIN-INAF 2010 grant.
We acknowledge the CINECA Award N. HP10CA62TX, 2012 for the availability of high-performance computing resources and support. 
We gratefully thank the bwGRiD project for the computational resources.
A.M. wishes to thank A. Tevzadze for helpful comments on the hydrodynamics vortex problem, G. Mamatsashvili for kindly providing us the reference solution for the MHD shearing wave test, and G. Bodo for valuable comments.
\end{acknowledgements}

% for the bibliography, at the end
\bibliographystyle{aa} % style aa.bst
\bibliography{mignone.bib} % your references Yourfile.bib

%%%%%%%%%%%%%%%%%%%%%%%%%%%%%%%%%%%%%%%%%%%%%%%%%%%%%%%%%%%%%%%%%%%%%%%%%%
\appendix %First online appendix
\section{Equations in Different Systems of Coordinates}
\label{sec:coord_eqns}
%
%
%
%
%
%
%%%%%%%%%%%%%%%%%%%%%%%%%%%%%%%%%%%%%%%%%%%%%%%%%%%%%%%%%%%%%%%%%%%%%%%%%%

In this section, we give explicit expressions for the MHD equations (\ref{eq:mhd1_rho}) through (\ref{eq:mhd1_B}) written in terms of the residual velocity $\vec{v}'=\vec{v}-\vec{w}$ for different systems of coordinates.
Also, to make notations more compact, we use $\vec{m} = \rho\vec{v}$ and $\vec{m}'=\rho\vec{v}'$ to denote the total and residual momentum and $\vec{\E}'= -\vec{v}'\times\vec{B}$ will be adopted as a short-hand notation for the (residual) electric field.

%%%%%%%%%%%%%%%%%%%%%%%%%%%%%%%%%%%%%%%%%%%%%%%%%%%%%%
\subsection{Cartesian coordinates}
%
%%%%%%%%%%%%%%%%%%%%%%%%%%%%%%%%%%%%%%%%%%%%%%%%%%%%%%

In Cartesian coordinates $(x,y,z)$, we assume, without any loss of generality, that the bulk orbital motion takes place along the $y$ direction, i.e., $\vec{w}=w\hvec{y}$.
The magnitude of $w$ can be defined either analytically or by averaging in the $y$ direction. In any case, $w\equiv w(x,z)$ so that $\partial_y w = 0$.
Writing the equations explicitly
\begin{equation}\begin{array}{lcl}
\DS\pd{\rho}{t} + \nabla\cdot\left(\rho\vec{v}'\right) 
                + w\pd{\rho}{y}  & = & 0 \\ \noalign{\medskip}
\DS\pd{m_x}{t} + \nabla\cdot\left(m_x\vec{v}' - B_x\vec{B}\right) 
               + \pd{p_t}{x} 
               + w\pd{m_x}{y} & = & S_{m_x} \\ \noalign{\medskip}
\DS\pd{m_y'}{t} + \nabla\cdot\left(m_y'\vec{v}' - B_y\vec{B}\right) 
                + \pd{p_t}{y} 
                + w\pd{m_y'}{y} & = & S_{m'_y} \\ \noalign{\medskip}
\DS\pd{m_z}{t} + \nabla\cdot\left(m_z\vec{v}' - B_z\vec{B}\right) 
               + \pd{p_t}{z} 
               + w\pd{m_z}{y} & = & S_{m_z} \\ \noalign{\medskip}
\DS\pd{E'}{t} + \nabla\cdot\Big[(E'+p_t)\vec{v}' - \vec{B}\left(\vec{v}'\cdot\vec{B}\right)\Big]
              + w\pd{E'}{y} &=& S_{E'} \\ \noalign{\medskip}
\DS\pd{B_x}{t} + \pd{\E'_z}{y} - \pd{\E_y}{z}
               + w\pd{B_x}{y} & = & 0 \\ \noalign{\medskip}
\DS\pd{B_y}{t} + \pd{\E'_x}{z}  - \pd{\E'_z}{x} 
               - \pd{(wB_x)}{x} - \pd{(wB_z)}{z}  & = & 0 \\ \noalign{\medskip}
\DS\pd{B_z}{t} + \pd{\E_y}{x} - \pd{\E'_x}{y} 
               + w\pd{B_z}{y}  & = & 0 \,.
\end{array}\end{equation}
The source terms on the right-hand side of the momentum and energy equations may be written using the reduced forms given in Eq. (\ref{eq:Sm0}) and (\ref{eq:SE0}).
In this case, one has
\begin{equation}\begin{array}{lcl}
S_{m_x}  &=& \DS -\rho\pd{\Phi}{x}  \\ \noalign{\medskip}
S_{m'_y} &=& \DS -\rho\pd{\Phi}{y} - \rho\vec{v}'\cdot\nabla w \\ \noalign{\medskip}
S_{m_z}  &=& \DS -\rho\pd{\Phi}{z} \\ \noalign{\medskip}
S_{E'} &=&  -\rho\vec{v}'\cdot\nabla\Phi 
       - \rho v'_y\left(\vec{v}\cdot\nabla w\right)
       + B_y\left(\vec{B}\cdot\nabla w\right)\,.
\end{array}\end{equation}
However, following the guidelines given in \S\ref{sec:source_terms}, a more convenient discretization is given by using Eq. (\ref{eq:Sm1}) and (\ref{eq:SE1}).
This affects only the source terms appearing in the $y$-component of the momentum equation and in the energy equation which can now be cast as
\begin{equation}\begin{array}{lcl}
S_{m'_y} &=&\DS  -\rho\pd{\Phi}{y} - \nabla\cdot(w\rho\vec{v}') 
               + w\nabla\cdot(\rho\vec{v}') \\ \noalign{\medskip}
S_{E'}      &=& -\rho\vec{v}'\cdot\nabla\Phi
       + w\nabla^r\cdot(m'_y\vec{v}'-B_\phi\vec{B}) 
       + w\nabla^r\cdot(\rho w\vec{v}') + \\ \noalign{\medskip}
       & &\DS - \frac{w^2}{2}\nabla\cdot(\rho\vec{v}') 
              - \nabla\cdot\left[\frac{\rho w^2}{2}\vec{v}'
              - w\left(m'_y\vec{v}'-B_\phi\vec{B}\right)\right] \,.
\end{array}\end{equation}
%

%%%%%%%%%%%%%%%%%%%%%%%%%%%%%%%%%%%%%%%%
\subsubsection{Shearing-box equations}
\label{sec:SB}
%
%%%%%%%%%%%%%%%%%%%%%%%%%%%%%%%%%%%%%%%%

As a particular case, we briefly review the equations of the shearing-box model introduced in \S\ref{sec:shearingbox}.
By adopting a non-inertial frame that co-rotates with the disk at orbital frequency $\Omega_0$, the momentum and energy equations (\ref{eq:mhd_mom} and \ref{eq:mhd_E}) become, respectively,
\begin{equation}\begin{array}{lcl}
\label{eq:SB-mhd_mom+en}
\DS
\pd{(\rho\vec{v})}{t} + \nabla\cdot\left(\rho\vec{v}\vec{v} 
    - \vec{B}\vec{B}\right)
 + \nabla p_t &=& \rho\vec{g}_{s} - 2\Omega_0\hvec{z}\times\rho\vec{v}
\\ \noalign{\medskip}
\DS
\pd{E}{t} + \nabla\cdot\left[\left(E+p_t\right)\vec{v} 
                           - \left(\vec{v}\cdot\vec{B}\right)\vec{B}\right]
 &=& \rho\vec{v}\cdot\vec{g}_{s} \,,
\end{array}
\end{equation}
where $\vec{g}_{s} = \Omega_0^2(2qx\hvec{x} - z\hvec{z})$ is the tidal expansion of the effective gravity while $q$ is the shear parameter (Eq \ref{eq:SB_q}) and the second term in Eq. (\ref{eq:SB-mhd_mom+en}) represents the Coriolis force.
The continuity and induction equations retain the same form as the original system in Eq. (\ref{eq:mhd_rho}) and (\ref{eq:mhd_B}).

The derivation of Eq. (\ref{eq:SB-mhd_mom+en}) with FARGO-MHD is done similarly to the previous section with $\vec{w} = -q\Omega_0 x\hvec{y}$ leading to
\begin{equation}\label{eq:SB-mhd1_mom+E}
\begin{array}{lcl}
\DS \pd{(\rho\vec{v}')}{t} + \nabla\cdot\left(\rho\vec{v}'\vec{v}' 
    - \vec{B}\vec{B}\right)
 + \nabla p_t &=& \vec{S}_{m'}
\\ \noalign{\medskip}
\DS \pd{E'}{t} + \nabla\cdot\left[\left(E'+p_t\right)\vec{v}' 
            - \left(\vec{v}'\cdot\vec{B}\right)\vec{B}\right]
 &=& S_{E'} \,,
\end{array}
\end{equation}
where the source term $\vec{S}_{m'}$ and $S_{E'}$ can be shown to be, respectively, equal to
\begin{eqnarray}
 \vec{S}_{m'} &=& 
 2\rho\Omega_0v'_y\hvec{x} + \rho\Omega_0(q-2)v_x\hvec{y} 
                           - \rho\Omega_0^2z\hvec{z}
\\ \noalign{\medskip}
 S_{E'} &=& -\rho v_z\Omega_0^2z - \left(B_yB_x-\rho v'_yv_x\right)q\Omega_0
\,.
\end{eqnarray}
We note that only the vertical component of gravity is included in the orbital frame \citep{StoGar.2010}. 
%

%%%%%%%%%%%%%%%%%%%%%%%%%%%%%%%%%%%%%%%%%%%%%%%%%%%%%%
\subsection{Polar coordinates}
\label{sec:polcoords}
%
%%%%%%%%%%%%%%%%%%%%%%%%%%%%%%%%%%%%%%%%%%%%%%%%%%%%%%

In polar cylindrical coordinates $(R,\phi,z)$, we consider the bulk orbital velocity to be aligned with the azimuthal direction, that is, $\vec{w} = \Omega R\hvec{\phi}$ where $\Omega=\Omega(R,z)$ is the angular rotation velocity defined by averaging $v_\phi/R$ along the azimuthal direction. 
Writing the equations in components yields
\begin{equation}\begin{array}{lcl}
\DS\pd{\rho}{t} + \nabla\cdot\left(\rho\vec{v}'\right) 
                + \Omega\pd{\rho}{\phi}  & = & 0 \\ \noalign{\medskip}
\DS\pd{m_R}{t} + \nabla\cdot\left(m_R\vec{v}' - B_R\vec{B}\right) 
               + \pd{p_t}{R} 
               + \Omega\pd{m_R}{\phi} & = & S_{m_R} \\ \noalign{\medskip}
\DS\pd{m_\phi'}{t} + \nabla^R\cdot\left(m_\phi'\vec{v}' - B_\phi\vec{B}\right) 
                   + \frac{1}{r}\pd{p_t}{\phi} 
                   + \Omega\pd{m_\phi'}{\phi} & = & S_{m'_\phi} \\ \noalign{\medskip}
\DS\pd{m_z}{t} + \nabla\cdot\left(m_z\vec{v}' - B_z\vec{B}\right) 
               + \pd{p_t}{z} 
               + \Omega\pd{m_z}{\phi} & = & S_{m_z} \\ \noalign{\medskip}
\DS\pd{E'}{t} + \nabla\cdot\Big[(E'+p_t)\vec{v}'  
              - \vec{B}\left(\vec{v}'\cdot\vec{B}\right)\Big]
              + \Omega\pd{E'}{\phi} &=& S_{E'} \\ \noalign{\medskip}
\DS\pd{B_R}{t} + \frac{1}{R}\pd{\E'_z}{\phi}
               - \pd{\E_\phi}{z}
               + \Omega\pd{B_R}{\phi} & = & 0  \\ \noalign{\medskip}
\DS\pd{B_\phi}{t} + \pd{\E'_R}{z}  - \pd{\E'_z}{R} 
                  - \pd{(wB_R)}{R} - \pd{(wB_z)}{z} & = & 0 \\ \noalign{\medskip}
\DS\pd{B_z}{t} + \frac{1}{R}\pd{(R\E_\phi)}{R} 
               - \frac{1}{R}\pd{\E'_R}{\phi} 
               + \Omega\pd{B_z}{\phi}  & = & 0 \,,
\end{array}
\end{equation}
where, besides the usual divergence operator $\nabla\cdot(\,)$, we have introduced the ``augmented'' divergence
\begin{equation}
 \nabla^R\cdot\vec{F} = \frac{1}{R^2}\pd{(R^2F_R)}{R} 
                          + \frac{1}{R}\pd{F_\phi}{\phi}
                          + \pd{F_z}{z}  \,,
\end{equation}
which avoids the curvature source terms in the $\phi-$component of the residual momentum equation, and is more appropriate for ensuring total angular-momentum conservation.

The source terms in the momentum and energy equations can be written, using the reduced form Eq. (\ref{eq:Sm0}) and (\ref{eq:SE0}), as
\begin{equation}\begin{array}{l}
S_{m_R} = \DS  -\rho\pd{\Phi}{R} 
             + \frac{\rho v_\phi^2 - B_\phi^2}{R} \\ \noalign{\medskip}
S_{m'_\phi} =\DS -\frac{\rho}{R}\pd{\Phi}{\phi}
               - \rho\vec{v}'\cdot\nabla w - \frac{\rho v_Rw}{R} 
               \\ \noalign{\medskip}
S_{m_z} =\DS -\rho\pd{\Phi}{z} \\ \noalign{\medskip}
S_{E'}  =\DS -\rho\vec{v}'\cdot\nabla\Phi
        - \left(  \rho v'_\phi \vec{v}- B_\phi \vec{B}\right)\cdot\nabla w 
        + \frac{\rho v_Rv_\phi w-B_RB_\phi w}{R} 
\end{array}
\end{equation}
We note that the total velocity $v_\phi$ appears in the radial momentum source term.
To restore full conservation of total angular momentum and total energy, the formulation given by Eq. (\ref{eq:Sm1}) and (\ref{eq:SE1}) is more appropriate. 
This leads to 
\begin{equation}\begin{array}{lcl}
S_{m'_\phi} &=&\DS  -\frac{\rho}{R}\pd{\Phi}{\phi}
               - \nabla^R\cdot(w\rho\vec{v}')
               + w\nabla\cdot(\rho\vec{v}') \\ \noalign{\medskip}
S_{E'}   &=&\DS  -\rho\vec{v}'\cdot\nabla\Phi
            + w\nabla^R\cdot(m'_\phi\vec{v}'-B_\phi\vec{B}) 
            + w\nabla^R\cdot(\rho w\vec{v}') + \\ \noalign{\medskip}
    & & \DS - \frac{w^2}{2}\nabla\cdot(\rho\vec{v}') 
            - \nabla\cdot\left[\frac{\rho w^2}{2}\vec{v}'
            - w\left(m'_\phi\vec{v}'-B_\phi\vec{B}\right)\right]\,.
\end{array}
\end{equation}

%%%%%%%%%%%%%%%%%%%%%%%%%%%%%%%%%%%%%%%%%%%%%%%%%%%%%%
\subsection{Spherical coordinates}
\label{sec:sphcoords}
%
%%%%%%%%%%%%%%%%%%%%%%%%%%%%%%%%%%%%%%%%%%%%%%%%%%%%%%

In spherical coordinates $(r,\theta,\phi)$ the direction of orbital motion coincides with the azimuthal direction, i.e., $\vec{w} = \Omega r\sin\theta\hvec{\phi}$ where $\Omega\equiv\Omega(r,\theta)$ is total angular rotation velocity computed by averaging $v_\phi/(r\sin\theta)$ in the $\phi$ direction. 
Writing equations (\ref{eq:mhd1_rho}) through (\ref{eq:mhd1_B}) in components yields
\begin{equation}\begin{array}{lcl}
\DS\pd{\rho}{t} + \nabla\cdot\left(\rho\vec{v}'\right) 
             + \Omega\pd{\rho}{\phi}  & = & 0 \\ \noalign{\medskip}
\DS\pd{m_r}{t} + \nabla\cdot\left(m_r\vec{v}' - B_r\vec{B}\right) 
            + \pd{p_t}{r} + \Omega\pd{m_r}{\phi} & = & S_{m_r} \\ \noalign{\medskip}
\DS\pd{m_\theta}{t} + \nabla\cdot\left(m_\theta\vec{v}' - B_\theta\vec{B}\right) 
                    + \frac{1}{r}\pd{p_t}{\theta} 
                    + \Omega\pd{m_\theta}{\phi} & = & S_{m_\theta} \\ \noalign{\medskip}
\DS\pd{m'_\phi}{t} + \nabla^r\cdot\left(m_\phi'\vec{v}' - B_\phi\vec{B}\right) 
                   + \frac{1}{r\sin\theta}\pd{p_t}{\phi} 
                   + \Omega\pd{m_\phi'}{\phi} & = & S_{m'_\phi} \\ \noalign{\medskip}
\DS\pd{E'}{t} + \nabla\cdot\Big[(E'+p_t)\vec{v}'  
           - \vec{B}\left(\vec{v}'\cdot\vec{B}\right)\Big]
           + \Omega\pd{E'}{\phi} &=& S_{E'} \\ \noalign{\medskip}
\DS\pd{B_r}{t} + \frac{1}{r\sin\theta}\pd{(\sin\theta\E_\phi)}{\theta} 
               - \frac{1}{r\sin\theta}\pd{\E'_\theta}{\phi} 
               + \Omega\pd{B_r}{\phi} &=& 0 \\ \noalign{\medskip}
\DS\pd{B_\theta}{t} + \frac{1}{r\sin\theta}\pd{\E'_r}{\phi} 
                    - \frac{1}{r}\pd{(r\E'_\phi)}{r} 
                    + \Omega\pd{B_\theta}{\phi} &=& 0 \\ \noalign{\medskip}
\DS\pd{B_\phi}{t} + \frac{1}{r}\pd{(r\E'_\theta)}{r} 
                  - \frac{1}{r}\pd{\E'_r}{\theta} 
                  - \frac{1}{r}\pd{(rwB_r)}{r} 
                  - \frac{1}{r}\pd{(wB_\theta)}{\theta} &=& 0 
\end{array}
\end{equation}
where, in analogy with the polar system of coordinates, we have introduced
the ``augmented'' divergence operator
\begin{equation}\begin{array}{lcl}
 \nabla^r\cdot\vec{F}&=&\DS \frac{1}{r^3}\pd{}{r}(r^3F_r) 
     + \frac{1}{r\sin^2\theta}\pd{}{\theta} (\sin^2\theta F_\theta)
\\ \noalign{\medskip}
& & \DS
     + \frac{1}{r\sin\theta}\pd{F_\phi}{\phi}  \,,
\end{array} 
\end{equation}
which is more convenient for expressing conservation of total angular momentum.

The source terms in the momentum and energy equations can be written, using the reduced form Eq. (\ref{eq:Sm0}) and (\ref{eq:SE0}), as
\begin{equation}\begin{array}{lcl}
S_{m_r} &=&\DS  -\rho\pd{\Phi}{r}
            + \frac{\rho v_\theta^2 - B_\theta^2}{r}
            + \frac{\rho v_\phi^2 - B_\phi^2}{r}\\ \noalign{\medskip}
S_{m_\theta} &=&\DS -\frac{\rho}{r}\pd{\phi}{\theta}
                - \frac{\rho v_\theta v_r - B_\theta B_r}{r}
                +\cot\theta \frac{\rho v_\phi^2 - B_\phi^2}{r}\\ \noalign{\medskip}
S_{m'_\phi} &=&\DS -\frac{\rho}{r\sin\theta}\pd{\Phi}{\phi}
                - \rho\vec{v}\cdot\nabla w 
                - \frac{\rho w}{r}\left(v_r+\cot\theta v_\theta\right)  \\ \noalign{\medskip}
S_{E'} &=&\DS -\rho\vec{v}'\cdot\nabla\Phi
       + \frac{(\rho v_r v_\phi - B_rB_\phi)w}{r} 
\\ \noalign{\medskip}
    & & \DS
       + \cot\theta\frac{(\rho v_\theta v_\phi - B_\theta B_\phi) w}{r}
       - \left(\rho v'_\phi \vec{v} - B_\phi\vec{B}\right)\cdot\nabla w\,.
\end{array}
\end{equation}
Again, in order to enforce conservation of both total angular momentum and total energy, the formulation given by Eq. (\ref{eq:Sm1}) and (\ref{eq:SE1}) is the more appropriate to adopt. 
In this case, one rewrites the $\phi$-momentum and energy source terms as
\begin{equation}\begin{array}{lcl}
S_{m'_\phi} &=&\DS -\frac{\rho}{r\sin\theta}\pd{\Phi}{\phi}
                - \nabla^r\cdot(w\rho\vec{v}')
                + w\nabla\cdot(\rho\vec{v}') \\ \noalign{\medskip}
S_{E'} &=&\DS -\rho\vec{v}'\cdot\nabla\Phi
       + w\nabla^r\cdot(m'_\phi\vec{v}'-B_\phi\vec{B}) 
       + w\nabla^r\cdot(\rho w\vec{v}') +
\\ \noalign{\medskip}
    & &\DS - \frac{w^2}{2}\nabla\cdot(\rho\vec{v}') 
           - \nabla\cdot\left[\frac{\rho w^2}{2}\vec{v}'
           - w\left(m'_\phi\vec{v}'-B_\phi\vec{B}\right)\right]\,.
\end{array}
\end{equation}

\end{document}